\documentclass[fleqn,usenatbib]{mnras}

\usepackage{newtxtext,newtxmath}
\usepackage{hyperref}
\usepackage{academicons}
\usepackage{aas_macros}
\usepackage{soul}
\usepackage{amsmath}
\usepackage{tikz}
\usepackage{graphicx}
\usepackage[T1]{fontenc}

\newcommand{\orcid}[1]{\href{https://orcid.org/#1}{\textcolor[HTML]{A6CE39}{\aiOrcid}}}

\def\hi{H\,{\sc i} }

\DeclareRobustCommand{\VAN}[3]{#2}
\let\VANthebibliography\thebibliography
\def\thebibliography{\DeclareRobustCommand{\VAN}[3]{##3}\VANthebibliography}

\title[Predictions for Detecting the bTFR Turndown]{Predictions for Detecting a Turndown in the Baryonic Tully Fisher Relation}

\author[Ruan et al.]{
Dilys Ruan,$^{1}$\thanks{E-mail: druan@physics.rutgers.edu}
Alyson M. Brooks,$^{1}$
Akaxia Cruz$^{2,3}$,
Annika H. G. Peter$^{4,5,6}$,
Benjamin W. Keller$^{7}$,
\newauthor
Thomas Quinn$^{8}$,
James Wadsley$^{9}$,
and Elizabeth A. K. Adams$^{10,11}$ \\
$^{1}$ Department of Physics and Astronomy, Rutgers University, Piscataway, NJ 08854, USA \\
$^{2}$ Center for Computational Astrophysics, Flatiron Institute, 162 Fifth Avenue, New York, NY 10010, USA \\
$^{3}$ Department of Physics, Princeton University, Princeton, NJ 08544, USA \\ 
$^{4}$ CCAPP, The Ohio State University, 191 W Woodruff Ave, Columbus, OH 43210, USA\\
$^{5}$ Department of Physics, The Ohio State University, 191 W Woodruff Ave, Columbus, OH 43210, USA\\
$^{6}$ Department of Astronomy, The Ohio State University, 140 W 18th Ave, Columbus, OH 43210, USA \\ 
$^{7}$ Department of Physics and Materials Science, The University of Memphis, Memphis, TN 38152, USA\\
$^{8}$ Astronomy Department, University of Washington, Seattle, WA 98195, USA \\
$^{9}$ Department of Physics and Astronomy, McMaster University, Hamilton, L8S 4M1, Canada \\
$^{10}$ ASTRON, the Netherlands Institute for Radio Astronomy, Oude
Hoogeveensedijk 4,7991 PD Dwingeloo, the Netherlands \\
$^{11}$ Kapteyn Astronomical Institute, University of Groningen, PO Box 800, 9700 AV Groningen, the Netherlands
}

\date{Accepted XXX. Received YYY; in original form ZZZ}

\pubyear{2025}

\begin{document}

\label{firstpage}
\pagerange{\pageref{firstpage}--\pageref{lastpage}}
\maketitle

\begin{abstract}
    The baryonic Tully Fisher relation (bTFR) provides an empirical connection between baryonic mass and dynamical mass (measured by the maximum rotation velocity) for galaxies.  Due to the impact of baryonic feedback in the shallower potential wells of dwarf galaxies, the bTFR is predicted to turn down at low masses from the extrapolated power-law relation at high masses. The low-mass end of the bTFR is poorly constrained due to small samples and difficulty in connecting the galaxy’s gas kinematics to its dark matter halo. Simulations can help us understand this connection and interpret observations. We measure the bTFR with 75 dwarf galaxies from the Marvel-ous and Marvelous Massive Dwarfs hydrodynamic simulations. Our sample has M$_\star = 10^6-10^9$ M$_\odot$, and is mostly gas dominated. We compare five velocity methods: V$_\text{out,circ}$ (spatially resolved mass-enclosed), V$_\text{out,mid}$ (spatially resolved midplane gravitational potential), and unresolved \hi linewidths at different percentages of the peak flux (W$_\text{10}$, W$_\text{20}$, and W$_\text{50}$). We find an intrinsic turndown in the bTFR for maximum halo speeds $\lesssim 50$ km s$^{-1}$, or total baryonic mass M$_\text{bary}\lesssim 10^{8.5}$ M$_\odot$. We find that observing \hi in lower-mass galaxies to the conventional surface density limit of 1 M$_\odot$ pc$^{-2}$ is not enough to detect a turndown in the bTFR; none of the \hi velocity methods, spatially resolved or unresolved, recover the turndown, and we find bTFR slopes consistent with observations of higher-mass galaxies.  However, we predict that the turndown can be recovered by resolved rotation curves if the \hi limit is $\lesssim 0.08$ M$_\odot$ pc$^{-2}$, which is within the sensitivity of current \hi surveys like FEASTS and MHONGOOSE.
\end{abstract}


\section{Introduction}

Rotation curves provided some of the earliest evidence for dark matter halos \citep{1978Bosma, 1980Rubin} since rotation velocities trace the dynamical mass. The Tully Fisher Relation (TFR) \citep{1977TullyFisher} is a measurement of galaxy stellar mass (or luminosity) as a function of rotation velocity, and thus provides an empirical constraint between a galaxy's stellar mass and underlying dark matter halo mass. Similar to the stellar-mass halo-mass (SMHM) relation \citep[e.g.,][]{Wechsler2018, 2021ApJ...923...35M}, the TFR helps us understand the galaxy-halo connection and galaxy evolution. Both the TFR and SMHM relation can tell us what fraction of the cosmic baryon abundance has cooled into halos and formed stars.  The baryonic Tully Fisher Relation (bTFR) instead connects baryonic mass (M$_\text{bary}$, total cold gas + stellar mass) to halo mass \citep{2000McGaugh}. This relation is especially useful to study lower-mass galaxies, in which the baryonic content is dominated by cold gas (T $<10^4$ K) in the atomic hydrogen (\hi) phase.
   
For galaxies with M$_\text{bary} \sim 10^8 - 10^{12}$ M$_\odot$, baryonic mass and the rotation velocity of gas follow a power-law with a bTFR slope of $3-4$ \citep[e.g.,][]{2009Stark,2010Gurovich,2016LelliBTFR, 2016BradfordBTFR, 2019Lelli}. The bTFR slope demonstrates a similar scaling between the baryonic disks and dark matter halos of galaxies, despite the many processes which redistribute baryons and dark matter \citep[e.g.,][]{2012Pontzen, 2014DiCintio}. The value of the slope has also been proposed to test different models of gravity, where a slope $= 4$ is consistent with modified Newtonian dynamics \citep{1983Milgrom,2012McGaugh}. Scatter in the bTFR tells us how much this scaling differs from galaxy-to-galaxy. Intrinsic scatter in the bTFR is $\sim 0.15$ dex, and is mostly due to scatter in the mass-concentration relation between halos \citep{2012Dutton}. 

For lower-mass galaxies (M$_\text{bary}\lesssim 10^8$ M$_\odot$), the bTFR is expected to steepen, or have a `turndown', because baryonic feedback leads to an overall decrease in star formation efficiency and loss of gas \citep[e.g.,][]{2012Ferrero, Oman2016}. These processes can be internal to the galaxy (e.g., galactic winds resulting from stellar feedback), or these processes can be external (e.g., UV background radiation from post-reionization, which can heat gas and prevent accretion). Hydrodynamic simulations predict a turndown in the bTFR around a rotation speed of 40 km s$^{-1}$ if the maximum rotation velocities of the dwarf galaxy halos are used \citep{2012Ferrero, 2016Brook, 2017Sales, 2018trujillogomez, 2024Sardone}. 

Measuring the bTFR at lower masses is difficult due to limited sample sizes of low-surface brightness galaxies and difficulty in matching gas to halo kinematics. Only a handful of resolved rotation curves have been measured in dwarf galaxies \citep[e.g.,][]{2015Oh}. Although \hi gas extends farther out in radius than stars \citep[e.g., ][]{1984Krumm, 1997BandRhee, 2008Begum, 2012Hunter,2024Hunter}, the \hi may not extend far out enough to where the galaxy rotates at its maximum speed. Many measurements to date extend out to a radius where the \hi surface density drops below 1 M$_{\odot}$ pc$^{-2}$ \citep[e.g.,][]{1997BandRhee,2016Wang}. At this limit, some observed rotation curves are `still rising', suggesting that the maximum rotation velocity is not being directly measured. 
\citet{2022McQuinn} constrained the bTFR with 25 galaxies at M$_\text{bary}< 10^8$ M$_\odot$. They used the outermost \hi velocities, and found no turndown in the bTFR.  However, using an adopted mass model, they were able to extrapolate the maximum rotation velocities that best matched each of their outermost rotation velocities. Assuming a cored Einasto dark matter density profile, the extrapolated velocities yielded a bTFR turndown at $\sim 40$ km s$^{-1}$ and baryon fractions of 1-10\% of the cosmic value. 
    
Current and upcoming radio surveys such as Apertif \citep{2022vanCapellen,2022Adams}; FEASTS \citep[FAST Extended Atlas of Selected Targets Survey,][]{Wang2025}; MHONGOOSE \citep[MeerKAT \hi Observations of Nearby Galactic Objects: Observing Southern Emitters, ][]{2024deBlok}; MIGHTEE-\hi \citep[MeerKAT International GigaHertz Tiered Extragalactice Exploration survey - HI,][]{2021Maddox}; and WALLABY \citep[Widefield ASKAP L-band Legacy All-sky Blind surveY,][]{2020Koribalski, 2020Spekkens, 2024Deg} are sensitive down to expected \hi masses of galaxies with rotation speeds of $\sim30-40$ km s$^{-1}$, below where the bTFR turndown is expected. However, these resolved rotation curves will still be limited to the nearest dwarf galaxies. The time required to obtain large sample sizes will still be prohibitive. For representative samples, spatially \textit{unresolved} \hi velocities are thought to be necessary, since they require less integration time and can be measured by both radio interferometers and single-dish telescopes. Studies with unresolved \hi velocities \citep[e.g.,][]{2015Klypin,2016LelliBTFR,2016BradfordBTFR} can have about two orders of magnitude more galaxies than spatially resolved observations \citep[e.g.,][]{2015Klypin,2016PapastergisTBTF,2022McQuinn}. \citet{2024Sardone} demonstrated that spatially unresolved, high spectral resolution observations are needed to detect the most diffuse gas in the outer wings of the \hi profile, and this can reduce bTFR scatter at lower masses. 

It is well established that different velocity tracers of the maximum halo speed (V$_\text{max}$) exhibit different bTFR slopes. This has been studied in observations \citep[e.g.,][]{2001Verheijen, 2016BradfordBTFR, 2019Lelli} and simulations \citep[e.g.,][]{2016Brook_diffv, 2017Brooks, 2020Glowacki, 2024Sardone}. For spatially unresolved \hi velocities, linewidths at 50\% (W$_{50}$) or 20\% (W$_{20}$) of the peak flux emission are reported as proxies for 2V$_\text{max}$. Higher-mass galaxies have flat rotation curves, and therefore more material with high column densities moving at the same velocity V$_\text{max}$. This high-velocity gas leads to a `double-horned' spectral feature that makes W$_\text{50}$ a better proxy to 2V$_\text{max}$. In the case of lower-mass galaxies, where \hi rotation curves are still rising and the contribution from non-circular motions is greater \citep{ElBadry2018}, the \hi flux profile is more Gaussian, and W$_{50}$ underestimates 2V$_\text{max}$ \citep[e.g.,][]{2016Brook_diffv, 2016Yaryura}. W$_{20}$ can provide a more accurate tracer of 2V$_\text{max}$ than W$_{50}$ \citep{2016Maccio, 2016Brook,  2017Brooks, 2019Dutton, 2024Sardone}, though still deviates at lower masses. Although W$_{10}$ (linewidth at 10\% of the peak \hi flux) is the largest linewidth, and therefore closest to 2V$_{\rm max}$ for dwarf galaxies, it is also susceptible to noise from outer velocity channels. 

We utilize the largest suite of simulated dwarf galaxies, with realistic \hi content, to understand how the bTFR turndown might manifest in observations. In this work, we analyze our sample of 75 dwarf galaxies from the Marvel-ous and Marvelous Massive Dwarfs hydrodynamic simulation suites. Our sample consists of isolated galaxies with M$_\text{HI} > 10^6$ M$_\odot$, and M$_\star = 10^6 - 10^9$ M$_\odot$. We focus on understanding why \hi velocities (spatially resolved or unresolved) do not trace the halo's V$_\text{max}$. We also make predictions about whether the halo's V$_\text{max}$ is recoverable with deeper \hi observations. In Section \ref{sims_methods}, we describe the properties and physics implemented in our simulations, and define the velocities and masses used in our study. Section \ref{results} presents our results and demonstrates how velocity definitions and sensitivity limits affect our ability to measure a bTFR turndown. Section \ref{discussion} discusses how our bTFR results compare to observations and upcoming \hi surveys. Section \ref{conclusions} summarizes our results. 

\section{Simulations \& Methods}
    \label{sims_methods}
    High resolution ``zoom-in'' simulations allow us to take a larger volume, identify high-density regions of interest (e.g., a halo of a given mass), and embed higher resolution particles in a smaller region to simulate the star formation and baryonic feedback with better spatial and temporal resolution. The large volume at lower resolution (external to the high-resolution particles) is needed because large-scale structure is responsible for the build-up of angular momentum in the main halo, according to tidal torque theory \citep{1969Peebles, 1987Barnes}.  The zoom-in is run with higher resolution and hydrodynamics for a region on the order of a few virial radii from the main halo's center.

    Halos are identified with \textsc{Amiga's Halo Finder} (AHF) \citep{2004MNRAS.351..399G, 2009ApJS..182..608K}. We define the virial radius of a halo as the distance from the center of mass in which the average halo density is 200 times that of the critical density at a given redshift. Much of the analysis is performed using \textsc{Pynbody} \footnote{\url{https://pynbody.github.io/pynbody/}} \citep{2013ascl.soft05002P}, a python package for SPH + N-body simulations.

    In this work, we use two smoothed particle hydrodynamics (SPH) + N-body zoom-in simulation sets, the Marvel-ous and Marvelous Massive Dwarfs suites (hereafter, Marvel and Massive Dwarfs, respectively). Marvel \citep{2021ApJ...923...35M} is a set of four zoom-in regions that each contain dozens of dwarf galaxies. The simulations are called CptMarvel, Elektra, Rogue, and Storm. The other simulation suite, Massive Dwarfs, consists of zoom-ins of individual halos selected from the Romulus25 simulation \citep{2017MNRAS.470.1121T}, focusing on dwarf galaxies with M$_\star\sim10^{8-9}$ M$_{\odot}$.

    Large-scale structure depends on the input cosmology. The baryon content will also be influenced by the cosmic fraction, which is set by the matter density and baryon density parameters, $\Omega_m$ and $\Omega_b$, respectively.  We assume $\Lambda$CDM and do not consider alternative types of dark matter. Marvel uses a WMAP3 cosmology \citep{2007ApJS..170..377S} ($h=0.732$, $\Omega_m=0.276$, $\Omega_bh^2=0.02229$). Massive Dwarfs uses the same cosmology as Romulus25, which is Planck 2015 \citep{2016A&A...594A..13P} ($h = 0.677$, $\Omega_m=0.307$, $\Omega_b h^2 =  0.02226$). Since our results mostly depend on the cosmic baryon fraction (WMAP3 with f$_b=0.1507$ versus Planck 2015 with f$_b=0.1582$), we do not expect that the different cosmologies will have a significant impact on our results.

    Due to the discreteness of particle simulations, forces can only be resolved to a minimum spatial scale, and each type of particle has a minimum mass. The Marvel simulation has a total low-resolution volume of (25 Mpc)$^3$ and uses 60 pc for the spline gravitational force resolution, 420 M$_\odot$ for initial star particle mass, 1410 M$_\odot$ for gas particle mass, and 6650 M$_\odot$ for dark matter particle mass. The Massive Dwarfs are run at a higher resolution than Romulus25 (3 times higher force resolution, 512 times higher mass resolution). Each  Massive Dwarfs simulation has a volume of (25 Mpc)$^3$, and uses 87 pc for spline gravitational force resolution, 994  M$_\odot$ for initial star particle mass, 3300  M$_\odot$ for gas particle mass, and 18000  M$_\odot$ for dark matter particle mass.

    Our simulations use the code \textsc{ChaNGa} \citep{2015ComAC...2....1M} for the SPH + N-body processes. The hydrodynamics are from \textsc{GASOLINE2} \citep{2004NewA....9..137W,2017MNRAS.471.2357W}, while \textsc{ChaNGa} has a faster gravity solver. \textsc{ChaNGa} scales well to simulations with a large number of particles due to its implementation in \textsc{Charm++} \citep{KaleKrishnan}, which handles dynamic load balancing and communication for thousands of computing cores. 

    \textsc{ChaNGa} calculates non-equilibrium ion abundances and gas cooling from collisional ionization rates \citep{1997Abel}, radiative recombination \citep{1981Black,1996Verner}, photo-ionization, free-free emission, and cooling from hydrogen and helium \citep{1992Cen}. The heating and cooling rates consider effects from a redshift-dependent, uniform UV background \citep{2012HaardyMadau}. Our simulations include the non-equilibrium formation, shielding, cooling, and destruction of H$_2$, as implemented by \citet{2012MNRAS.425.3058C}. Metal-line cooling and metal distribution are based on the model of  \citet{2010MNRAS.407.1581S}.

     We adopt the star formation probability from \citet{2006MNRAS.373.1074S} and require the presence of H$_2$ based on \citet{2012MNRAS.425.3058C}:
    \begin{equation}
         p = \frac{m_\text{gas}}{m_\text{star}} \left( 1 - e^{-c_0^{*} X_{H_2} \Delta t/t_\text{form}} \right)\text{,}
    \end{equation}
    where $m_\text{gas}$ is the gas particle mass, $m_\text{star}$ is the initial star particle mass, $c_0^*$ is the star formation efficiency parameter (set to 0.1), $X_{H_2}$ is the H$_2$ fraction of the gas particle, $\Delta t$ is the time step for star formation in the simulation ($\sim$1 Myr), and $t_\text{form}$ is the local dynamical time. Star formation is limited to cold gas particles ($T<1000$ K), and a number density $n>0.1$ cm$^{-3}$ is required. Since we also require the presence of H$_2$, star formation usually occurs in much denser gas, so the actual density for star formation is generally $n > 100$ cm$^{-3}$. Our star formation prescription from \citet{2012MNRAS.425.3058C} reproduces the Kennicut-Schmidt relation \citep{1959Schmidt,1998Kennicutt} -- e.g., see right panel of Figure 10 in \citet{Christensen2014}. In both Marvel and Massive Dwarfs, each star particle follows the initial mass function from \citet{2001Kroupa}.

    \textsc{ChaNGa} calculates the \hi mass fraction for every gas particle throughout the simulation. The gas fraction is based on the particle's temperature and density, heating from the cosmic UV background radiation and young stars, H$_2$ self-shielding, and dust shielding in \hi and H$_2$. We can use the \hi gas fractions and sum over particles to get the total \hi mass. Similarly, the total stellar mass and halo mass are calculated by summing over particles within the virial radius (R$_\text{200}$). \citet{2013Munshi} estimated that photometric observational methods yield stellar masses that are $\sim$60\% of simulation values, so we include this factor for a more direct comparison to observations. Throughout this study, any stellar mass will be 60\% of the simulation total. The total halo mass ($M_{200}$) uses the same virial constraint as \textsc{AHF} at $200\rho_c(z)$.

    Marvel uses blastwave feedback \citep{2006MNRAS.373.1074S}, while Massive Dwarfs uses superbubble feedback \citep{2014MNRAS.442.3013K}. Blastwave feedback injects metals, thermal energy, and mass into gas that is within the blastwave radius of the supernova (SN), as calculated by the blastwave solution for a SN remnant in \citet{1988blastwave}. The deposited energy is $\sim 1.5\times 10^{51}$ erg per SN. To prevent numerical overcooling, blastwave feedback relies on turning off radiative cooling in the surrounding gas, to limit thermal energy loss in supernovae for a time corresponding to the end of the snowplow phase. Superbubble feedback does not disable radiative cooling, but instead puts feedback-heated particles into a temporary two-phase state, with each phase cooling with a separate temperature and density.  These two states are kept in pressure equilibrium, and correspond roughly to the cold swept-up ISM and the hot SN-heated interior.  Mass flows from the cold phase to the hot via thermal evaporation \citep{1977Cowie} until the entire cold phase has been evaporated.  This evaporation process can continue by stochastically evaporating cool neighbors of hot, SN-heated gas particles.  Within resolved hot bubbles, temperatures gradients are smoothed by thermal conduction.  The superbubble model is better able to capture the effects of multiple SNe occurring over a range of timescales, drives stronger outflows, and produces a more realistic gas phase diagram than the earlier blastwave model. For superbubble feedback, energy is deposited at a rate of $1\times10^{51}$ erg per SN. Superbubble feedback is thought to have stronger outflows that expel low angular momentum gas and regulate star formation \citep{2021Mina}, yet \citet{2025Piacitelli} find no difference in the circumgalactic medium in a larger sample of galaxies combining the two feedback models.
    
    Although these simulations have different physics, we use galaxies from both in order to span a larger mass range. The Marvel-ous Dwarf volumes are dominated by lower-mass dwarfs, while the Massive Dwarfs simulations are (by definition) focused on higher stellar mass dwarfs of M$_\star\sim10^{8-9}$ M$_{\odot}$, and any additional field dwarf galaxies contained in the zoom region. We expect properties like mass, size, and rotation velocity to generally be consistent across simulations. \citet{2024arXiv240106041A} ran one set of the Marvel simulations with blastwave and superbubble feedback models, and demonstrated that both feedback models reproduced the observed trends for Local Volume dwarf galaxies in terms of their stellar mass, size, and metallicity. Galaxies with superbubble feedback exhibited higher halo masses, though the variance between feedback models was within the observational error of the SMHM relation. This result suggests that galaxies from the Massive Dwarfs simulation may have slightly higher rotation velocities in a given mass range (compared to galaxies from Marvel), but we expect the difference to be small.
 
   \citet{2025Piacitelli} also used galaxies from Marvel and Massive Dwarfs (with blastwave and superbubble feedback, respectively) and found consistent trends between simulations for specific star formation rate versus stellar mass, the luminosity-metallicity relation, and column density profiles for gas in the circumgalactic medium. Meanwhile, \citet{2021ApJ...923...35M} showed that the slightly different resolutions used in these simulation sets has no impact on the resulting SMHM relation.  In summary, there is no conclusive evidence that the differing feedback models or resolutions yield significantly different galaxy properties for the two suites.  Given these results, we combine them here for a sample that spans as large a mass range as possible.

\subsection{Sample Selection}
\begin{figure*}
         \centering
         \includegraphics[width=0.9\textwidth]{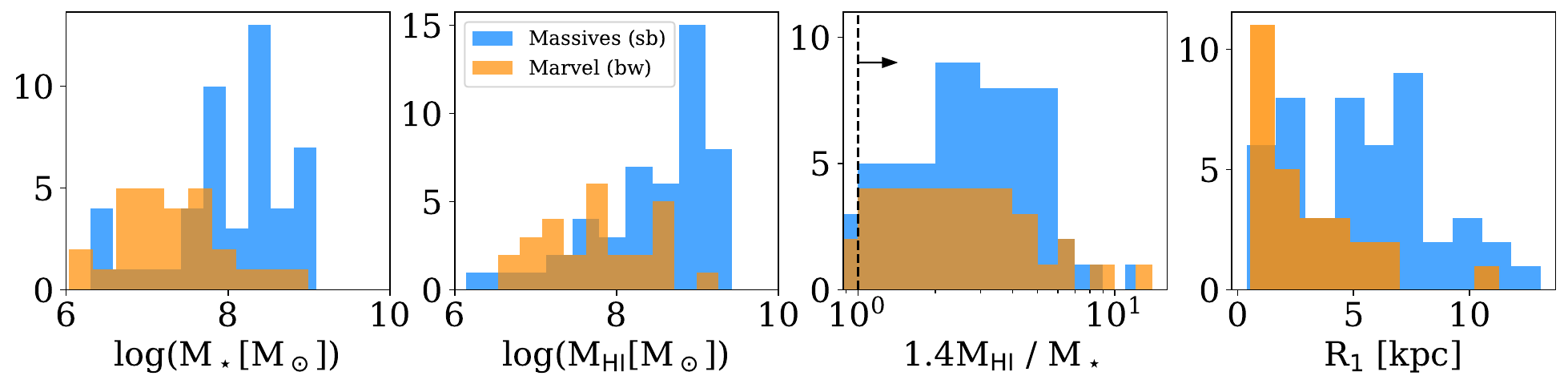}
         \caption{Histograms of (from left to right) stellar mass, \hi mass, gas to stellar mass ratio, and R$_\mathrm{1}$ (\hi size at the fiducial surface density limit of 1 M$_\odot$ pc$^{-2}$). Galaxies from the Marvel simulation (27 out of 75) are represented in orange, while galaxies from Massive Dwarfs (48 out of 75) are shown in blue. We use M$_\text{gas} = 1.4$ M$_\text{HI}$, which is commonly used in observations. Galaxies to the right of the black dashed vertical line are gas-dominated, which is most of our sample.}
         \label{gasdom}
     \end{figure*}

     The faintest galaxies in most bTFR studies have stellar masses on the order of $10^7$  M$_\odot$ \citep[e.g.,][]{2015ApJ...809..146B,2016Lelli}. In the case of \citet{2022McQuinn}, a few galaxies have stellar masses $\sim4\times10^6$  M$_\odot$. To better understand the bTFR turndown, our sample includes dwarf galaxies with M$_\star > 10^6$  M$_\odot$. We also require that each galaxy has M$_\text{HI} > 10^6$  M$_\odot$ in order to have enough gas for observable rotation speeds.
     We select isolated field galaxies to focus on the connection between gas and dark matter kinematics without significant complications from galaxy-galaxy interactions. \citet{2016BradfordBTFR} required that any galaxy in their bTFR sample with M$_\star<10^{9.5}$ M$_\odot$ be isolated, with the 2D projected distance to the nearest most massive host > 1.5 Mpc. All of our simulated galaxies have M$_\star<10^{9.5}$ M$_\odot$, and are only included if there are no other galaxies of equal or greater mass within the virial radius.
     Our sample includes 75 dwarf galaxies, of which 48 are from the Massive Dwarfs suite and 27 are from the Marvel suite. Our galaxy mass ranges are $\log$(M$_{\star}$[M$_\odot$]) = $6.03-9.09$, $\log$(M$_\text{bary}$[M$_\odot$])=$6.66-9.64$, and $\log$(M$_{200}$[M$_\odot$]) = $9.58-10.92$.  Mass and size distributions for our sample are shown in Figure \ref{gasdom}.  We define the total baryonic mass as M$_{\rm bary} = $ M$_{\star} + 1.4$M$_{\rm HI}$, where the factor of 1.4 accounts for helium \citep[e.g.,][]{1996Arnett,2016BradfordBTFR} and metals in the gas phase. Our galaxy sample consists of 5 stellar-dominated galaxies and 70 gas-dominated galaxies, as shown in Figure \ref{gasdom}. Using these simulations, \citet{2025Keith} found that galaxies with M$_\star < 10^8$ M$_\odot$ are typically more irregular while those at higher masses have disks. They determined this by using the principle axis ratios of the 3D ellipsoid that best fit the shape of the star particles, and the principle axes were calculated from the stellar moment of inertia tensor.

\subsection{Spatially unresolved \hi profiles}
    We use the program \textsc{TIPSY} \footnote{\url{https://github.com/N-BodyShop/tipsy}} (Katz \& Quinn 1994) to create mock \hi data cubes from our simulations. We use these data cubes to derive spatially unresolved velocity widths (W$_\text{10}$, W$_\text{20}$, and W$_\text{50}$, as defined in the next section). With \textsc{TIPSY}, we can visualize star, gas, and dark matter particles, and orient the halo at different inclinations. To make a data cube, we center on the halo of interest, subtract its center of mass velocity so that the profile is centered at 0 km s$^{-1}$, and orient it edge-on ($i = 90^\circ$) or at a random inclination. The disk is aligned based on the angular momentum of gas particles in a central region of high-density gas that is visually identified. The angular momenta for stars and dark matter are ignored.
    We orient the angular momentum vector around the $z$-axis of the simulation and align the galaxy according to this.
    
    After we give \textsc{TIPSY} a viewing angle, the code considers the line-of-sight velocity for each particle. Our simulations track the velocity of each particle according to Newton's equations of motion, and we have to separately consider velocity contributions from turbulence or thermal broadening. Dwarf galaxies can have velocity dispersions on the order of their rotation velocities, $\sim 10-15$ km s$^{-1}$ \citep[e.g.,][]{2009Tamburro}, and dispersions are likely driven by thermal velocities or SNe \citep{2009Tamburro, 2013StilpA,2013StilpB}. Our simulations have SNe injecting thermal energy into the interstellar medium. To account for thermal broadening in the \hi profiles, \textsc{TIPSY} `smears' the velocity contribution from each gas particle into a Gaussian centered on the line-of-sight velocity with a standard deviation of $\sigma = \sqrt{kT/(\mu m_p)}$, where $T$ is the gas particle's temperature, $m_p$ is the proton mass, and $\mu$ is the mean molecular weight of the ISM.  The mean molecular weight determines what the average particle mass is as a fraction of a proton mass, and this is calculated based on the metallicity and ionization state of the gas.
    
    We use a fixed velocity resolution of $2.6$ km s$^{-1}$ and measure each galaxy over the velocity range -200 to 200 km s$^{-1}$. This channel width is set to match the best spectral resolution in \citet{2015ApJ...809..146B}. Our data cubes have fixed spatial axes spanning from -100 to 100 kpc across each galaxy with a resolution of 3.7 kpc, though the spatial resolution has little impact on the unresolved linewidth results. We do not consider beam smearing effects or observational noise, but will include these effects in a future work. The emission spectra are converted from \hi mass to flux by inverting the relation
    \begin{equation}
        {M}_\mathrm{HI} = 2.356\times 10^5 {D}^2 \int S_v \mathrm{d}v\text{, }
    \end{equation}
    where $v$ is the velocity in km s$^{-1}$, $S_\nu$ is the spectrum in Jy beam$^{-1}$, and we set the distance $D = 3$ Mpc. The \hi profiles of our full sample are shown in Appendix \ref{fullsample}.

\subsection{Inclinations}
    \label{inclination}

Inclination corrections to observed velocities can be substantial at the dwarf galaxy scale.  As previously described, we orient both our disky and irregular galaxies such that the total angular momentum vector of the central gas lies along the $z$ direction, and then measure our \hi linewidths edge-on. Of course, this is not the same method employed in observations, where the projected elliptical shape of the galaxy is measured to determine an inclination. 
Usually, inclination angles are calculated based on the major and minor axes of the galaxy's stellar components from optical observations. Galaxies with inclinations lower than $i \lesssim 30^{\circ} -40^{\circ}$ are often not included in samples, since projection uncertainties are significant \citep[e.g.,][]{Oman2016, Read2016, 2016LelliBTFR, 2017Read, 2022McQuinn}.  Observations typically report inclination-corrected rotation velocities, and so we use the edge-on velocities regardless of galaxy morphology.

At M$_{\star} \lesssim 10^8$ M$_{\odot}$, our simulated dwarfs are predominantly irregular rather than disky \citep{2025Keith}.  For these lowest-mass dwarf galaxies, we have found that our orientation method does not always yield a result that is edge-on by visual inspection.  It is possible that there is no orientation that yields an apparent edge-on result, given their irregular nature.  Although the inclinations for these irregular galaxies are difficult to measure and correct for, the Gaussian \hi profiles of these lower-mass dwarf galaxies do not vary much by viewing angle (as quantified by the errors to the right of each panel in the bottom row of Figure \ref{BTFR}.  Our inclination procedure with simulations allows us to include dwarf galaxies that would otherwise be removed from observational samples.

Despite the differences in methodology from observations, we keep our full dwarf sample in order to quantify trends to as low mass as possible.  However, we highlight that systematic differences between our \hi linewidth results versus observations likely come from these unknown inclinations \textit{and} the inclusion of more irregular, low-mass galaxies. Throughout this work, we only show inclination-corrected linewidths (by using the edge-on results) and compare to inclination-corrected values from observations.

\subsection{Definitions for Velocities and Radii} 
    \begin{figure}
        \centering
        \begin{tikzpicture}
          \node (N1) at (0,0) {\includegraphics[width=0.45\textwidth]{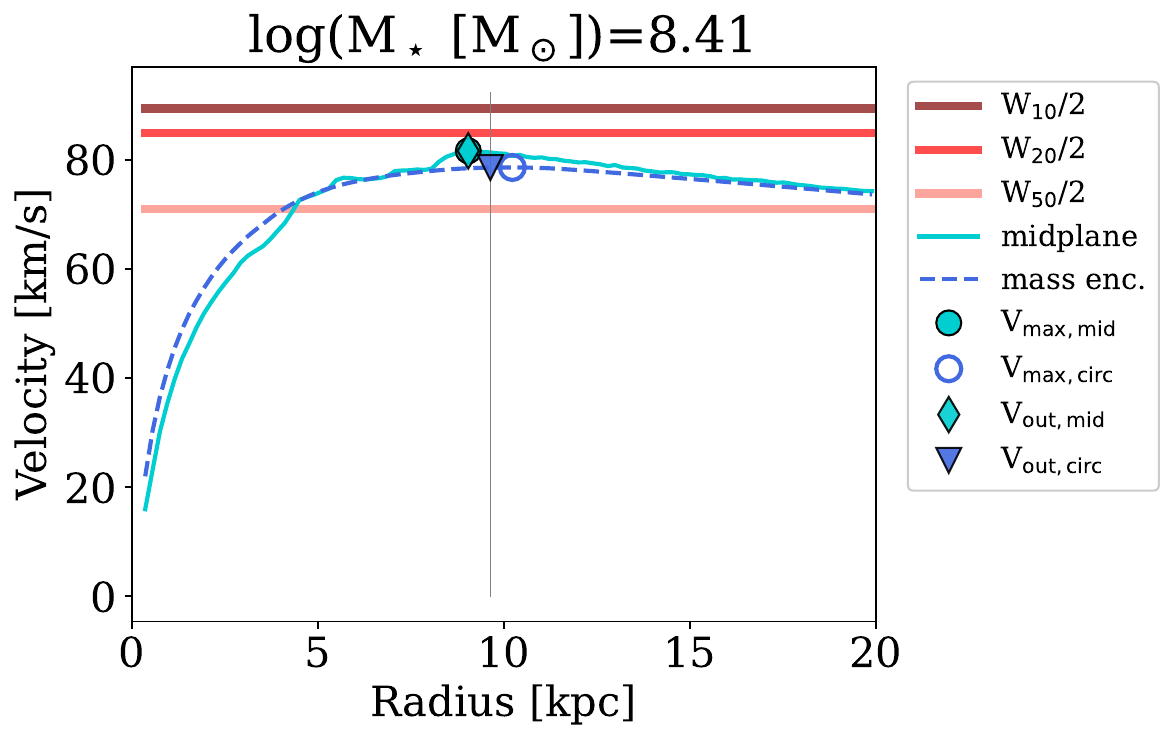}};
          \node[opacity=1.] (N2) at (0.,-0.4) {\frame{\includegraphics[width=0.17\textwidth]{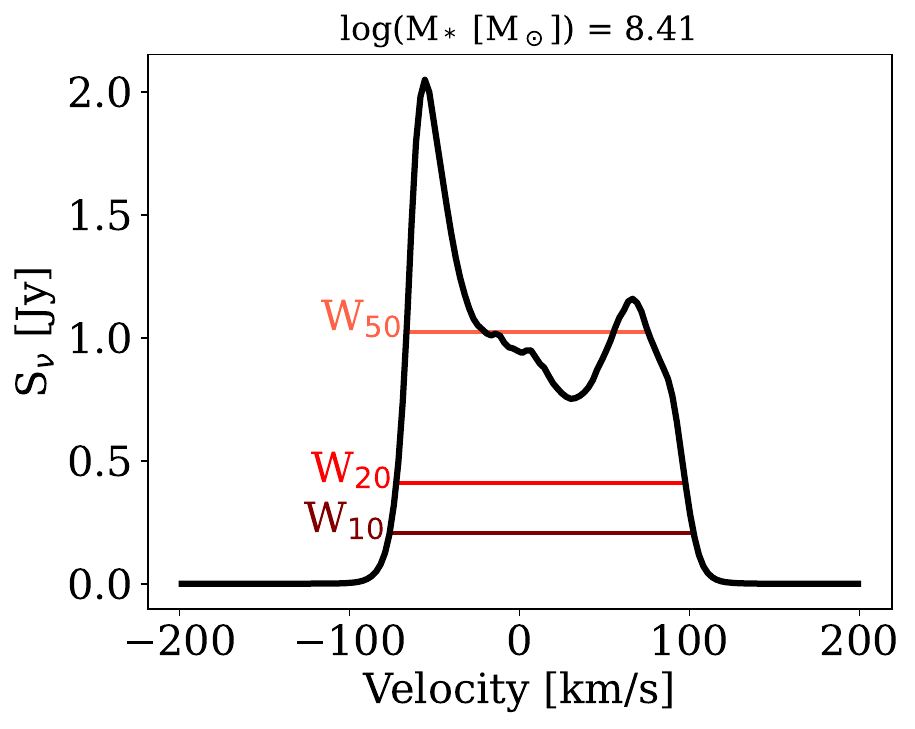}}};
        \end{tikzpicture}

         \begin{tikzpicture}
          \node (N1) at (0,0) {\includegraphics[width=0.45\textwidth]{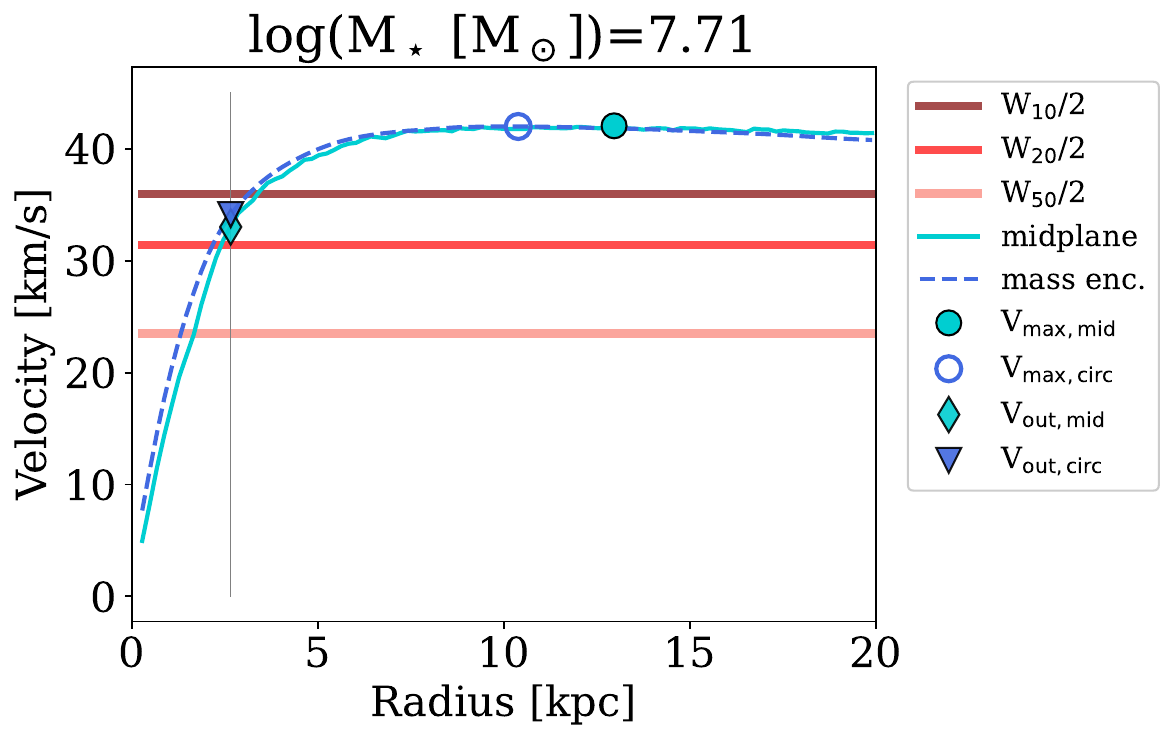}};
          \node[opacity=1.] (N2) at (0.,-0.4) {\frame{\includegraphics[width=0.17\textwidth]{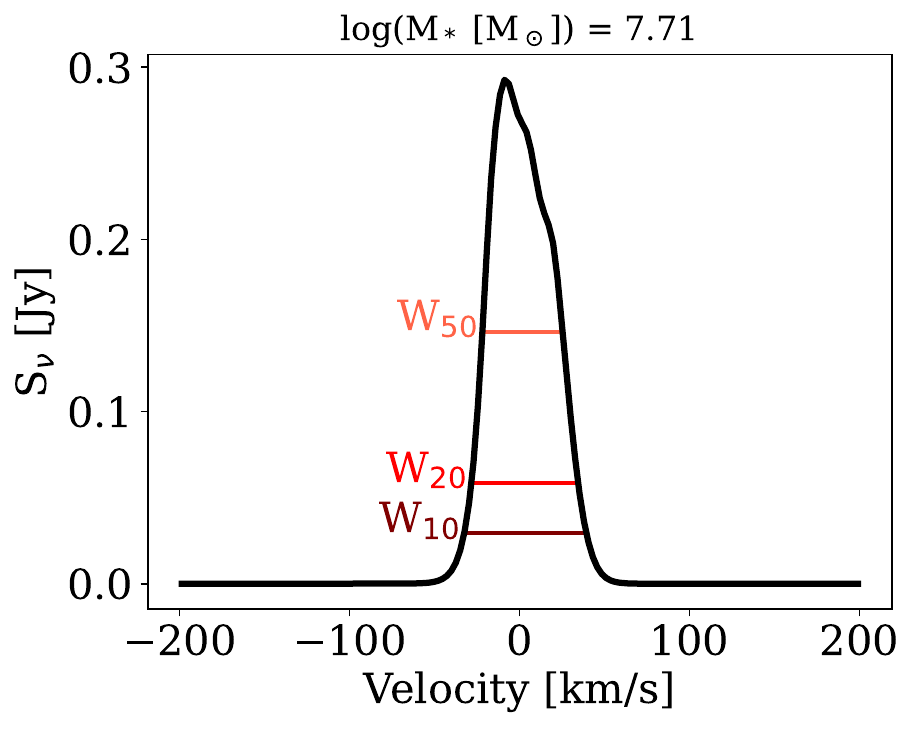}}};
        \end{tikzpicture}

        \caption{{Top panel} shows the \hi profile (inset) and rotation curve for r614 from the Massive Dwarfs suite, which is disky and has a higher stellar mass. {Bottom panel} is similar but for Rogue 8 from the Marvel suite, which is irregular and lower in mass. For the rotation curves, the solid cyan line is calculated using the midplane potential, whereas the dashed blue line uses the mass enclosed. The open blue point is V$_\text{max,circ}$, the solid cyan point is V$_\text{max,mid}$, the blue triangle represents V$_\text{out,circ}$, and the cyan diamond represents V$_\text{out,mid}$. The vertical grey line denotes the radius R$_1$, where the \hi surface density drops below 1 M$_\odot$ pc$^{-2}$.  \hi linewidths (W$_\text{x}$) are measured from the inset \hi profiles. Each linewidth is represented by a horizontal line of a different color (W$_\text{10}$/2 is maroon, W$_\text{20}$/2 is red, and W$_\text{50}$/2 is salmon). r614 has a higher stellar mass and exhibits the double-horned profile, while Rogue 8 has a more Gaussian profile. For the \hi emission profiles of our full sample, see Figure \ref{allhi}.}
    \label{rotationcurves}
    \end{figure}

    \begin{figure*}
        \centering
        \includegraphics[width=0.48\textwidth]{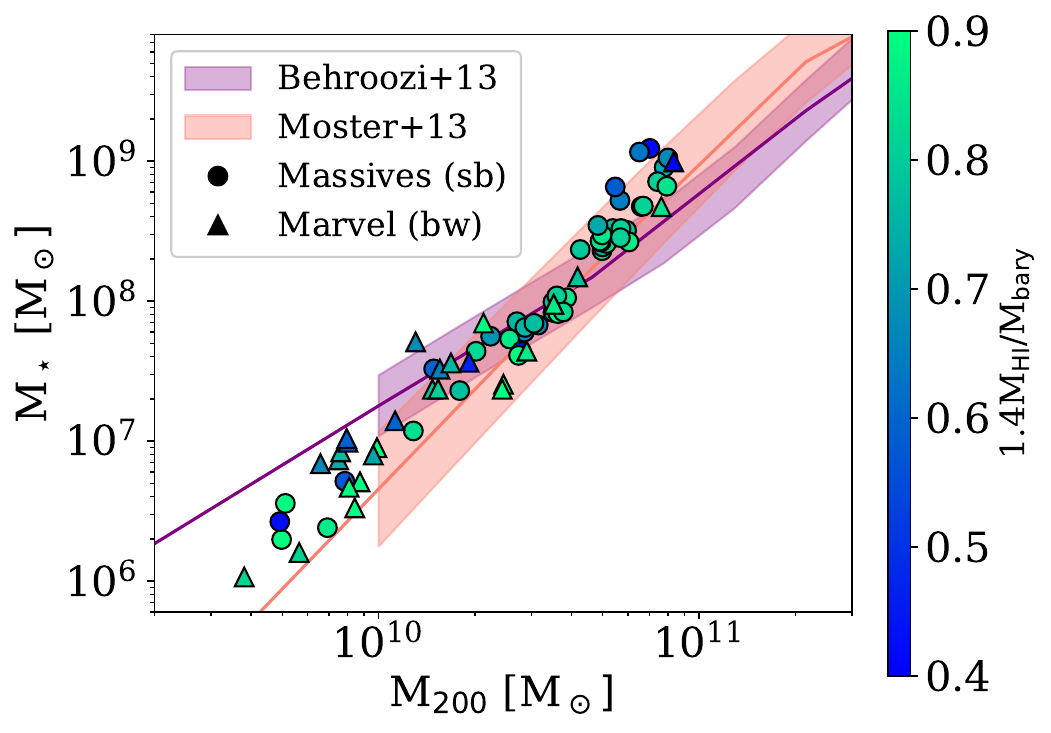}
        \includegraphics[width=0.48\textwidth]{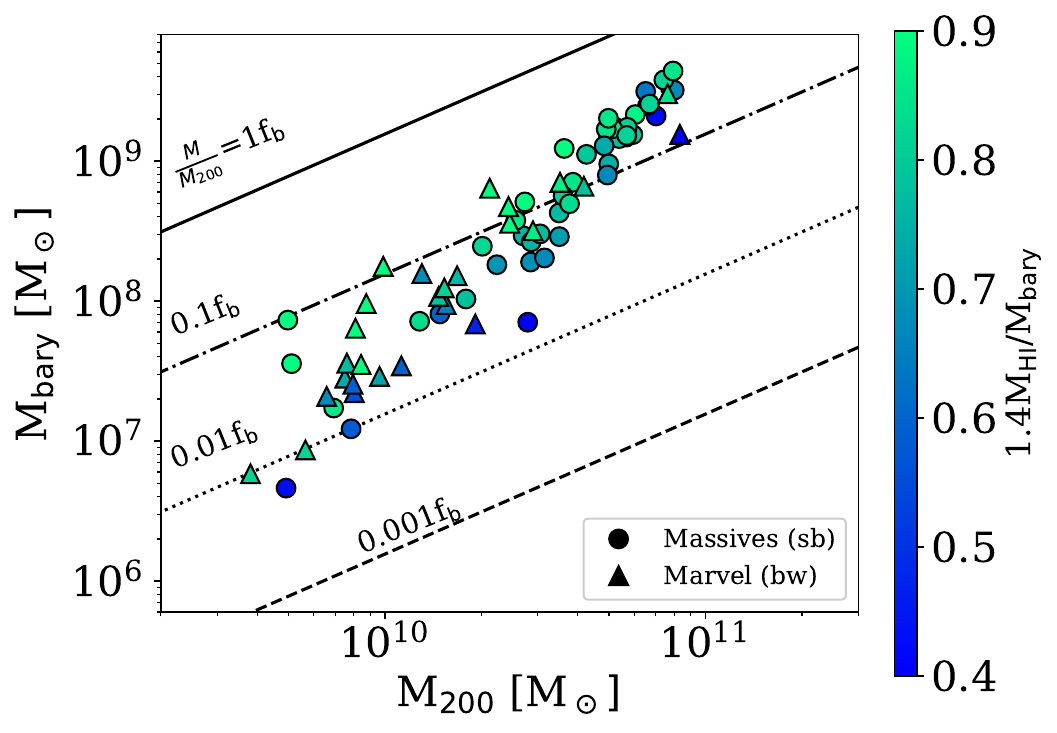}
        \caption{ 
        Left panel shows the SMHM relation, while the right panel shows the baryon-halo mass relation. In both panels, the Massive Dwarfs simulations that use superbubble (`sb') feedback are shown with points, while the dwarf galaxies from the Marvel simulations that use blastwave (`bw') feedback are shown with triangle markers. Each marker is colored by the gas fraction 1.4M$_\text{HI}$/M$_\text{bary}$ and shows the gas-rich nature of our sample where most of the galaxies have gas fractions $>$ 0.7. We compare our SMHM relation to abundance matching results from \citet{2013Behroozi} (purple shaded region) and \citet{2013Moster} (red shaded region) with 1-$\sigma$ error. Both abundance matching results were determined down to M$_\text{200}\sim10^{10}$  M$_\odot$, and we extrapolate the relations below this halo mass, indicated by the end of the shaded regions.
        Our stellar masses are 60\% of the total stellar mass calculated from the simulations, as this is a better match to photometrically-derived observational methods \citep{2013Munshi}. Our simulated galaxies are consistent with the SMHM constraints from abundance matching. We compare baryon mass fractions to the cosmic value (f$_\text{b} = \Omega_b / \Omega_m = 0.1556$) using cosmological values from \citet{2020Planck}). We plot M/M$_{200} = 1$f$_\text{b}$ (solid line), M/M$_{200} = 0.1$f$_\text{b}$ (dotted dashed line), M/M$_{200} = 0.01$f$_\text{b}$ (dotted line), and M/M$_{200} = 0.001$f$_\text{b}$ (dashed line). Both the SMHM and baryon-halo mass relations exhibit more scatter at M$_\mathrm{200} < 3\times10^{10}$  M$_\odot$.}
        \label{Masses}
    \end{figure*}
 
    There are a number of different ways to measure rotation velocity, and at different radii.  The methods for velocities and corresponding radii in this work are: 
    \begin{itemize}
        \item {\bf V$_\text{max,circ}$} and {\bf R$_\text{max,circ}$}: the maximum rotation velocity calculated based on concentric shells of the enclosed mass, as in the `dynamical mass', where the circular rotation velocity is defined as $V(R) = \sqrt{{GM(<R) }/{R}}$.  The radius corresponding to V$_\text{max,circ}$ is  R$_\text{max,circ}$.  The \textsc{Pynbody} profile class can calculate a rotation curve based on the mass enclosed. This is an idealized measurement enabled by the use of simulations. For our dwarf galaxies, R$_\text{max,circ}$ is often beyond the radius at which \hi could be observationally detected.
    
        \item {\bf V$_\text{max,mid}$} and {\bf R$_\text{max,mid}$}: the maximum rotation velocity calculated based on the midplane gravitational potential. R$_\text{max,mid}$ is the radius at which V$_\text{max,mid}$ occurs. The midplane rotation velocity follows the relation $V(R)  =  \sqrt{R \frac{\delta \Phi}{\delta R}}$, where the calculated velocity is based on the radial component of the gravitational acceleration of the entire galaxy using a few sample points in the midplane. \textsc{Pynbody} aligns the galaxy and determines this midplane according to the angular momenta of gas particles within 10 kpc from the halo's center. We orient the galaxy with the total gas angular momentum vector aligned with the $z$-axis, so that the disk is in the $x-y$ plane. Then, we calculate the rotation speed from this method as a function of radius with \textsc{Pynbody}'s profile class. Most of our galaxies are disky and align well in the $x-y$ plane.  However, we use this same methodology for all of the galaxies, even the non-disky ones where a disk plane is not well-defined, in order to calculate the midplane potential. 

        \item {\bf V$_\text{out,circ}$} and {\bf R$_\text{out,circ}$}:  R$_\text{1}$ is the 2D projected size of the \hi disk where its surface density drops below 1  M$_\odot$pc$^{-2}$ (corresponding to a column density of N$_\text{HI}\sim1.25\times10^{20}$ cm$^{-2}$). We set this value to match the canonical main \hi disk limit in other works \citep[e.g.,][]{1997BandRhee,2016Wang}. V$_\text{out,circ}$ is the maximum rotation speed using the mass-enclosed method at R $\leq$ R$_\text{1}$, and R$_\text{out,circ}$ is the corresponding radius.
        
        \item {\bf V$_\text{out,mid}$} and {\bf R$_\text{out,mid}$}: uses the same R$_\text{1}$ value, though the corresponding velocity and radius are determined from the rotation curve using the midplane gravitational potential.
     
        \item {\bf W$_\text{x}$ and R$_\text{x}$}: W$_\text{10}$, W$_\text{20}$, and W$_\text{50}$ are the \hi profile's width at different percentages of the peak flux height; 10\%, 20\%, and 50\%, respectively. Technically, the unresolved \hi linewidths are not set to a size extent, as the entire dwarf is usually detected within the single dish beam size. However, we calculate an equivalent radius by using the corresponding radius (R$_\text{x,circ}$) where the linewidth W$_\text{x}/2$ intersects the circular rotation curve. In some cases, W$_\text{x}/2$ does not intersect the rotation curve (see Figure \ref{rotationcurves}, top panel as an example). \hi linewidths may not exactly correspond to the circular velocity if the emission profile is asymmetric. If the gas is more extended on one side of the galaxy, hence moving at a faster speed, this results in a higher linewidth. Therefore, this linewidth can exceed the max velocity when assuming spherical shells of the same mass enclosed.

    \end{itemize}
     
     The \hi linewidths (W$_\text{10}$, W$_\text{20}$, and W$_\text{50}$) are spatially unresolved, while all other velocities are from rotation curves that are resolved as a function of radius. For the spatially resolved methods, we use linearly-spaced bins, starting at a radius of 3R$_\text{soft}$ and extending out to 20 kpc with 100 bins. The number of radial bins was selected based on the criteria from \cite{2003Power}, where $\gtrsim3000$ particles per enclosed region are needed for accurate circular velocities in dense central regions. R$_\text{soft}$ is the gravitational force softening length for each simulation suite: 0.087 kpc for the Massive Dwarfs and 0.060 kpc for Marvel. The spacing between radial bins is $\sim0.2$ kpc. 
     Figure \ref{rotationcurves} demonstrates how these velocities and radii compare for a higher-mass galaxy (r614) and lower-mass galaxy (Rogue 8) in our sample. r614 demonstrates how a higher-mass galaxy exhibits a flat rotation curve at the fiducial observational limit of 1 M$_\odot$ pc$^{-2}$, as well as a classic double-horned \hi profile. For a lower-mass galaxy like Rogue 8, the \hi gas does not extend as far out radially. Therefore, the galaxy's \hi velocities (spatially resolved or unresolved) do not match the galaxy's actual V$_\text{max,mid}$ or V$_\text{max,circ}$. 
     
      In this work, we aim to understand how the different methods and velocity definitions impact the determined bTFR.  We do not perform full mock observations, i.e., we do not include noise, beam smearing, or other observational effects in our \hi data cubes.  We also do not take into account the impact of non-circular motions, though we discuss the implications of this in Section \ref{sec:midplane}.  We assess whether the various velocity measurement methods have the potential to recover a turndown in the bTFR even with these `idealized' rotation curves and \hi linewidths.

     \label{definitions}

\section{Results}
\label{results}
\subsection{Comparing Stellar and \hi Properties to Observations}
     
    \begin{figure}
        \centering
        \includegraphics[width=0.45\textwidth]{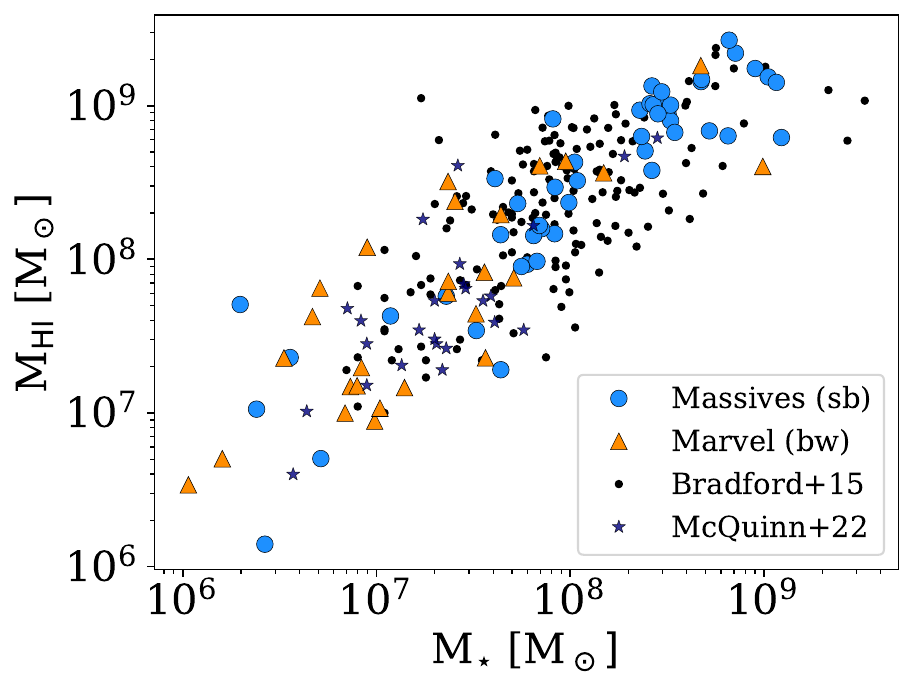}
        \includegraphics[width=0.475\textwidth]{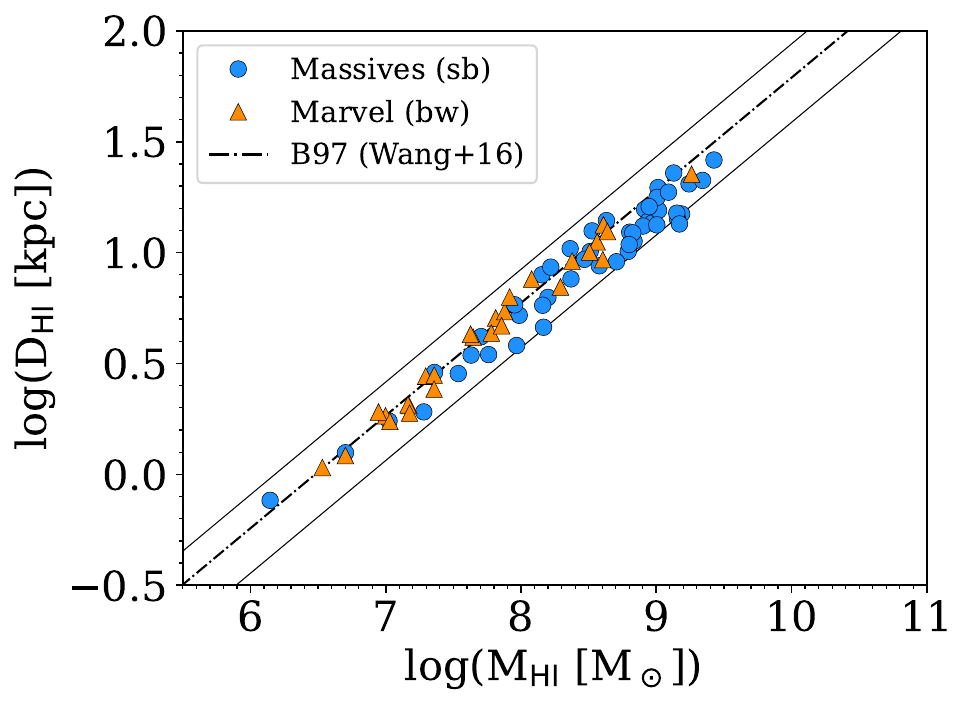}
        \caption{The top panel compares the total \hi mass as a function of stellar mass in our simulations (Massives - blue points, Marvel - orange triangles) to observations from \citet{2015ApJ...809..146B} (black points) and \citet{2022McQuinn} (navy stars). The bottom panel compares our simulations to the empirical relationship (dotted-dashed line) between the size and total mass of \hi gas in galaxies \citep{1997BandRhee}. In this study and the observations compared here, the \hi size is defined as the radius where the \hi surface density has dropped to 1  M$_\odot$ pc$^{-2}$. The solid black lines show the 3-$\sigma$ scatter in the observations from \citet{2016Wang}.}
        \label{HIsizemass}
    \end{figure}
    
    \begin{figure}
        \centering
        \includegraphics[width=0.45\textwidth]{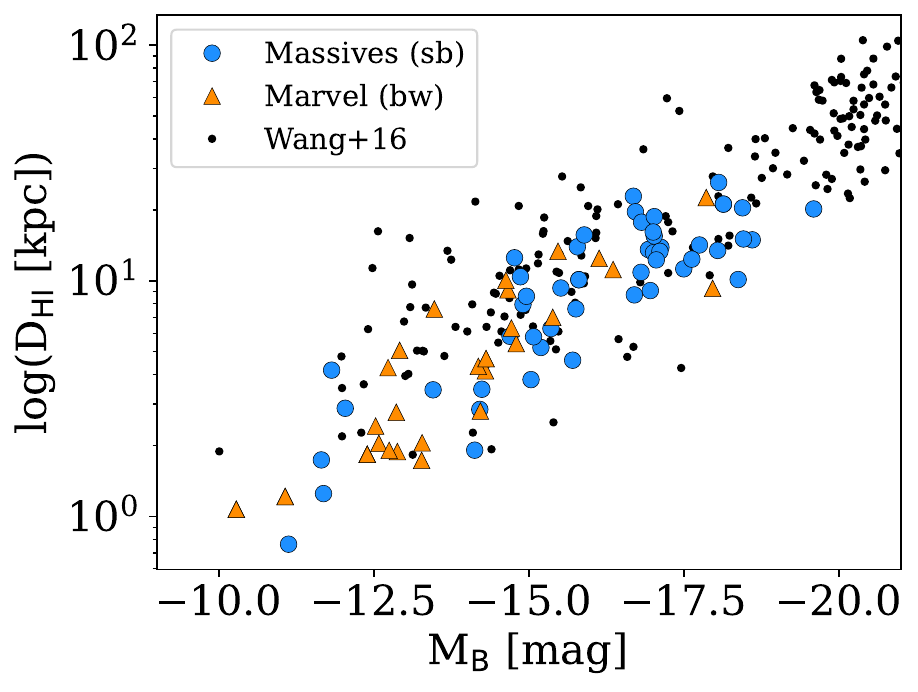}
        \includegraphics[width=0.45\textwidth]{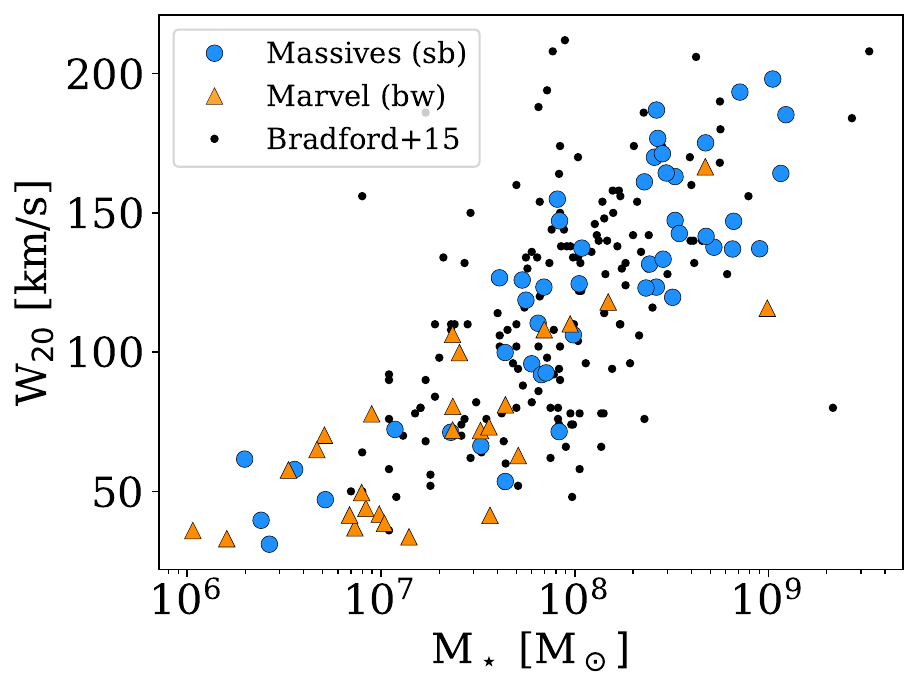}
        \caption{The top panel compares the \hi size to the stellar content as quantified by the absolute $B$-band magnitude. We calculate the cumulative $B$-band magnitude and compare our data (Massives - blue points, Marvel - orange triangles) to observations from \citet{2016Wang} (black points).  In general, we find the \hi properties of our simulated dwarf galaxies to be comparable to observed dwarf galaxies, though we may lack the most extended \hi disks. The bottom panel compares our linewidths from mock \hi data cubes to that of observed linewidths (black points) from \citet{2015ApJ...809..146B} for a given stellar mass. The observed linewidths are corrected for inclination and redshift broadening, and we plotted their values for 2V$_{20,i}$ to compare with our edge-on values of W$_\text{20}$, i.e., both the simulated and observed samples are inclination-corrected.}
        \label{sizes}
    \end{figure}
    
    Before analyzing the bTFR in our simulated dwarf galaxies, it is crucial to check whether we have achieved realistic baryonic properties for each galaxy. First, we check masses. Figure \ref{Masses} shows the SMHM relation for our sample in the left panel. We compare the SMHM relation from our simulations to the abundance matching results from \citet{2013Behroozi} and \citet{2013Moster}. For M$_\text{200} \gtrsim 3\times10^{10}$M$_\odot$, our SMHM relation matches \citet{2013Moster}, whereas at M$_\text{200} \lesssim 3\times10^{10}$ M$_\odot$, our data lie within the range of \citet{2013Behroozi} and \citet{2013Moster}. In the right panel of Figure \ref{Masses}, we show the baryon-halo mass relation for our sample. We define the total baryonic mass as M$_\text{bary}$ = M$_\star$ + 1.4M$_\text{HI}$, where the 1.4 factor also accounts for helium and metals in the gas phase.  We also compare our baryonic content to the cosmic fraction $f_b = \Omega_b/\Omega_m = 0.1582$ using the \citet{2020Planck} values: $\Omega_m = 0.315$, $\Omega_bh^2 = 0.0224$, and $h=0.674$. Our baryon fractions range from $\sim 0.01f_b - 0.5f_b$, and decrease with lower halo masses. Below M$_{200}\lesssim 3\times10^{10}$  M$_\odot$, we find more scatter in our SMHM relation and baryon-halo mass relation \citep[see also][]{2021ApJ...923...35M}.
   
    To assess whether our \hi mass content is realistic, we plot M$_\text{HI}$ as a function of M$_\star$ in the top panel of Figure \ref{HIsizemass}. We compare our simulated galaxies to the observed galaxies in \citet{2015ApJ...809..146B} and \citet{2022McQuinn}. \citet{2015ApJ...809..146B} combined data from their own observations, the ALFALFA survey \citep{2011Haynes}, and \citet{2006Geha}. In this comparison, we omit the ALFALFA data since \citet{2011Haynes} selected galaxies by \hi mass rather than stellar mass, which has been shown to be biased toward higher \hi masses than a stellar-mass selected sample \citep{2010Catinella}. 
    
    After confirming that our masses are consistent with observations, we check the \hi size-mass relationship in the bottom panel of Figure \ref{HIsizemass}. The size is defined as the diameter where the \hi surface density has dropped to 1  M$_\odot$ pc$^{-2}$, similar to observations. We compare the \hi size-mass relation from the simulated dwarf galaxies to the observations from \citet{2016Wang}, which follows the trend from \citet{1997BandRhee} (B97), and has been confirmed by recent WALLABY \citep{2024Deg} and FEASTS results \citep{Wang2025}. Our galaxies fit in the 3-$\sigma$ scatter of \citet{2016Wang}, represented by the solid black lines. \citet{2019Stevens} found that the \hi size-mass relation is robust to gas stripping and galaxy population (centrals versus satellites), so matching this relation mostly confirms that our feedback is of appropriate strength and does not disrupt the \hi disk. 

    Finally, because it has been suggested that galaxies simply evolve along the \hi size-mass relation \citep{2019Stevens}, we check whether our \hi disks extend to realistic radii by instead comparing them against stellar properties. In the top panel of Figure \ref{sizes}, we compare our \hi sizes against the absolute $B$-band magnitude (M$_\text{B}$). We compare the trend in our simulations to the observations in \citet{2016Wang}. We determine the galaxy's $B$-band brightness profile out to the virial radius with \textsc{Pynbody}'s profile class, which uses the \citet{2008Marigo} and \citet{2010Girardi} stellar population models. Then, we calculate the total B-band luminosity ($L$) by summing over radial bins, and turn this value into a magnitude with the relation M$_\text{B} = -2.5 \log_{10}(L*\alpha)$, where the conversion factor is $\alpha = 2.35 \times 10^9$ pc$^{2}$ arcsec$^{-2}$ erg$^{-1}$ s. Our galaxy sizes are within the observed range for M$_\text{B}=-12.5$ to $-18.0$, though we may not have examples of the most extended \hi disks.  This may be due to differences in our sample sizes and selection functions, since \hi surveys are more sensitive to galaxies with more \hi flux. 
    
    If our galaxies have \hi disks that are realistic in size, then their \hi linewidths should also match the observational range. In the bottom panel of Figure \ref{sizes}, we assess how well W$_\text{20}$ matches between simulations and observations at a given M$_\star$. \citet{2015ApJ...809..146B} corrected for inclination ($i$) and redshift ($z$) broadening in their linewidths, reporting the value V$_{20,i}$ = W$_\text{20} / \left[ 2\sin i (1+z)\right]$ as a maximum rotation velocity. Our edge-on linewidths should be comparable to the inclination-corrected velocities in \citet{2015ApJ...809..146B}, and so we plot their 2V$_{20,i}$ against our W$_\text{20}$. Our linewidths are indeed consistent with the range from \citet{2015ApJ...809..146B}, suggesting that our baryonic feedback is realistic, and so is the thermal broadening we used to make mock data cubes.

\subsection{bTFR with Different Velocity and Rotation Curve Definitions}
    \label{btfrmethodresults}
    Having shown that our simulations follow a number of observed scaling relations related to \hi, we now analyze the bTFR according to different velocity methods. 
    \begin{figure*}
        \centering
        \includegraphics[width=0.45\textwidth]{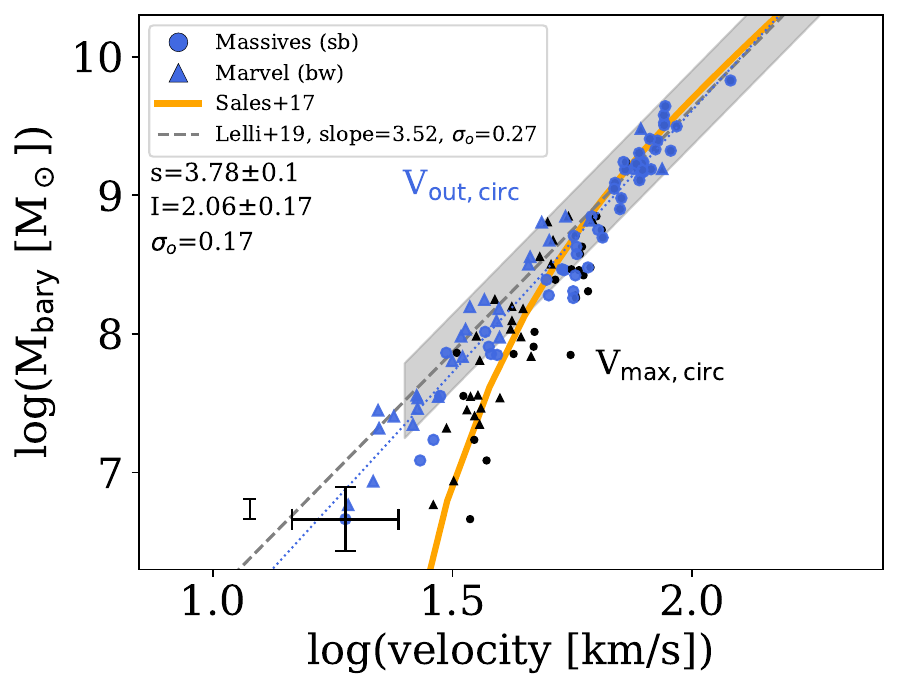}
        \includegraphics[width=0.45\textwidth]{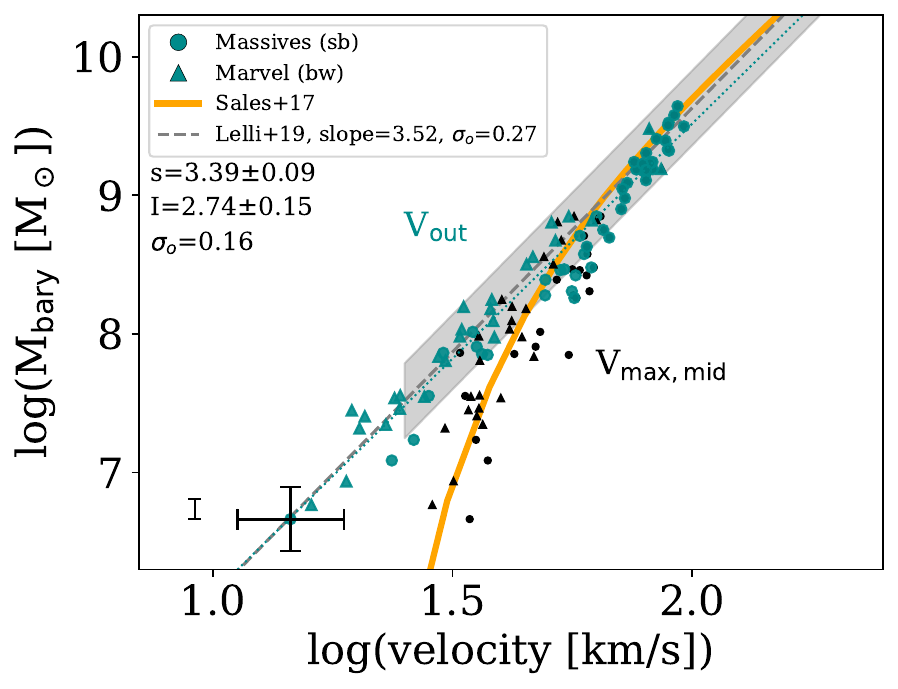}
        \includegraphics[width=0.45\textwidth]{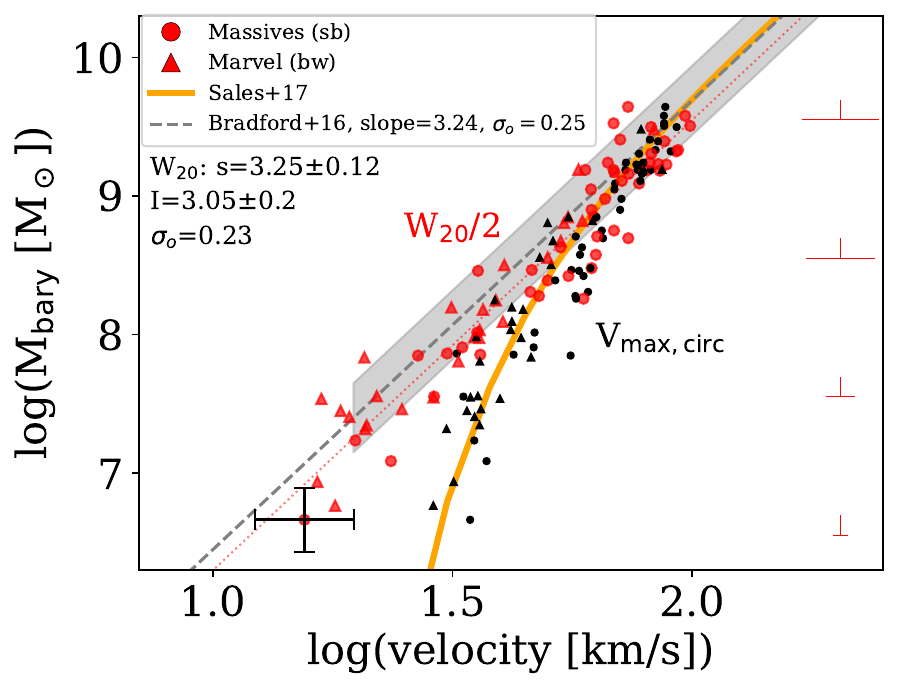}
        \includegraphics[width=0.45\textwidth]{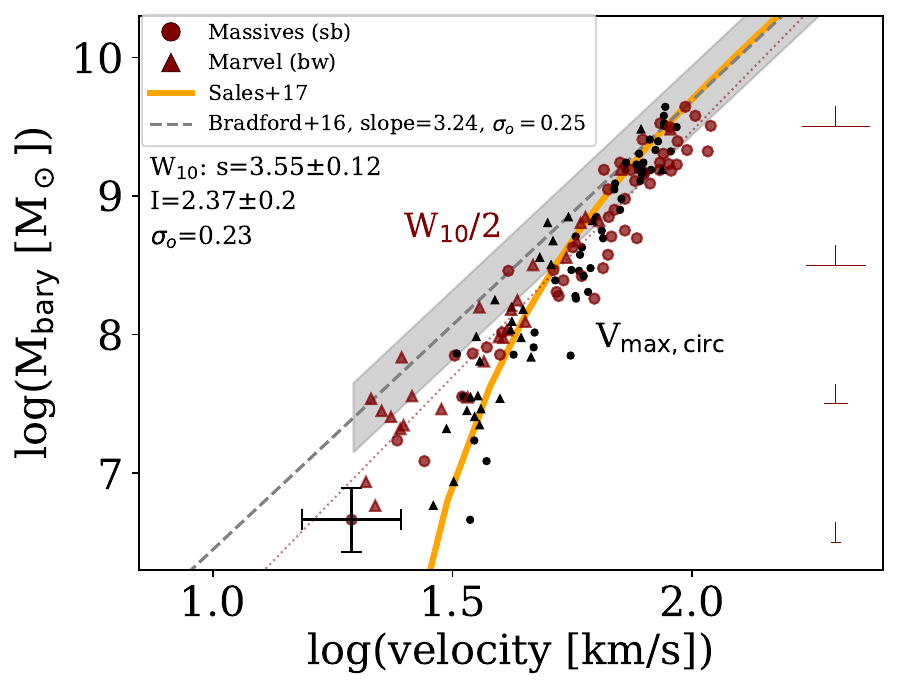}
        \caption{The bTFR from our simulations (markers: circles for Massives which uses superbubble (`sb') feedback, triangles for Marvel which uses blastwave (`bw') feedback), compared to constraints from observations in \citet{2019Lelli} (grey dashed line in the top panels) and \citet{2016BradfordBTFR} (grey dashed line in the bottom panels), as well as the APOSTLE simulations \citep[orange line,][]{2017Sales}. The black markers represent the max halo speed of each simulated galaxy, while the colored markers represent the same galaxy measured with \hi. Halo speeds are plotted as V$_\text{max,circ}$ (using the mass-enclosed potential) and V$_\text{max,mid}$ (using the midplane potential).  {Upper left panel} shows the bTFR when using V$_\text{out,circ}$ (spatially resolved mass-enclosed), upper right panel shows the bTFR with V$_\text{out,mid}$ (spatially resolved midplane potential), bottom left shows the bTFR when using the \hi linewidth W$_\text{20}/2$, while the bottom right uses W$_\text{10}/2$. Estimated observational errors are plotted on the lowest mass galaxy in each panel, and are derived in Appendix \ref{errorbar}. In the top panels, the mass error to the left of the observational errors shows the variance in stellar mass from simulations to observations according to \citet{2013Munshi}. In the bottom panels, simulation variance in stellar mass and linewidths (due to inclination angle) is plotted on the right side of each panel. We find a bTFR turndown using the \textit{halo's} maximum rotation speed, for either the mass-enclosed or gravitational midplane potential methods. However, none of our \hi gas methods trace the halo's maximum rotation speed at rotation speeds < 50 km s$^{-1}$ (or M$_\text{bary} \lesssim10^{8.5}$  M$_\odot$).}
        \label{BTFR}
    \end{figure*}

    \textbf{bTFR using the maximum halo speed:}
    Modeling the halo's circular velocity based on the total mass enclosed in spherical shells is common in bTFR studies \citep[e.g.,][]{2018trujillogomez,2022McQuinn,2019Lelli}, so we use this definition as our reference point. In Figure \ref{BTFR}, we show that the circular velocity of the halo (V$_\text{max,circ}$, black points) exhibits a turndown in the bTFR at M$_\text{bary} \lesssim 10^{8.5}$  M$_\odot$, corresponding to velocities $\lesssim$ 50 km s$^{-1}$, as galaxy formation becomes inefficient. Our bTFR trend is consistent with the turndown predicted by \citet{2017Sales} for the halo circular speed in the APOSTLE simulations (orange solid line). 
    
    We can also calculate the halo's rotation speed using the midplane gravitational potential (V$_\text{max,mid}$, black points -- top right panel in Figure \ref{BTFR}) instead of the enclosed mass. If galaxies are in dynamic equilibrium and the assumption of spherical symmetry is valid, the two methods should return the same velocity.  However, as explored in a number of previous works \citep[e.g.,][]{Read2016, ElBadry2017, Verbeke2017, Downing2023, Sands2024}, dwarf galaxies have bursty star formation histories with feedback that frequently drives them out of dynamical equilibrium.  This effect is likely to become more pronounced as galaxy mass decreases and the gravitational potential well becomes shallower and more susceptible to feedback.  Additionally, velocity dispersion becomes more dominant as galaxy mass decreases and rotation velocity declines, such that circular velocity is no longer an accurate measure of velocity.  Because of these effects, using the midplane gravitational potential may be a more accurate tracer of the underlying mass distribution, and perhaps more comparable to what an observer might measure at lower rotation speeds.  However, for our sample the overall  differences in the two methods are subtle (as seen in Figure \ref{rotationcurves}). We further discuss the effect of non-circular motions and the bTFR in Section \ref{sec:midplane}. Differences between the mass enclosed and midplane methods tend to be more significant at smaller radii, but our measurements for the maximum halo speed are measured at larger radii and hence both methods produce similar results for the bTFR. When using V$_\text{max,mid}$, we again find a turndown at M$_\text{bary} \lesssim 10^{8.5}$ M$_\odot$.

    As mentioned in Section \ref{sims_methods}, the superbubble feedback used in the Massive Dwarfs simulations (black circles) may result in slightly higher halo masses.  A close examination of the black points in Figure \ref{BTFR} suggests that the higher halo masses lead to slightly higher V$_\text{max}$ values in the Massive Dwarfs. In other words, the use of the combined simulations may broaden the scatter in the simulated bTFR.  Despite this, the variance in max halo speeds between feedback models is small, such that the spread introduced is still within the observational error. This is most obvious when comparing the V$_\text{max}$ spread in the \hi results (colored points, discussed in detail below) relative to the spread in observations (grey shaded regions).  The change due to feedback is small enough to not affect the conclusions of this work, and considering the effects of feedback on the bTFR turndown is beyond the scope of this paper.
    
    \textbf{bTFR using spatially resolved methods: } At an \hi surface density of 1 M$_\odot$ pc$^{-2}$, we measure the 2D projected radius, R$_\text{1}$. Then, we find the maximum rotation speeds for each rotation curve method at radii within R$_\text{1}$, which we call V$_\text{out,circ}$ (which uses the enclosed mass) and V$_\text{out,mid}$ (which uses the midplane potential). The top left panel of Figure \ref{BTFR} shows the bTFR when we use V$_\text{out,circ}$ (blue markers), while the top right panel shows the bTFR when using V$_\text{out,mid}$ (cyan markers). We compare our spatially resolved results to the V$_\text{max}$ trend in \citet{2019Lelli} (grey dashed line), which uses the maximum velocity from observed rotation curves, for galaxies that have both flat and rising rotation curves. From their Table 1, \citet{2019Lelli} measured a slope = 3.52, intercept = 2.59, and vertical scatter = 0.27 (grey shaded region) with a sample of 153 galaxies.
    Both spatially resolved \hi methods are consistent with observations as they each follow a power-law trend and do not exhibit a turndown. We show representative error bars on the lowest-mass galaxy, in which we estimate the observational errors for velocity and mass (discussed further in Appendix \ref{errorbar}). 

    \textbf{bTFR with \hi Linewidths: } In the bottom row of Figure \ref{BTFR}, we show the bTFR when using spatially unresolved \hi linewidths. We compare the linewidths (bottom left panel: W$_\text{20}$ in red markers, bottom right panel: W$_\text{10}$ in maroon markers) to V$_\text{max,circ}$ (black markers). The observational sample compared here is different than in the previous two panels.  The dashed line shows the bTFR result from \citet{2016BradfordBTFR}, which uses the W$_\text{20}$ linewidths for 930 isolated galaxies and has a slope = 3.24, intercept = 3.21, and vertical scatter = 0.25 (grey shaded region).  As established by other works \citep{2016Maccio, 2017Brooks, 2019Dutton,2024Sardone}, W$_\text{50}$ often underestimates the halo rotation speed (V$_\text{max,circ}$). W$_\text{20}$ and W$_\text{10}$ trace the halo at higher masses, but do not exhibit the bTFR turndown at lower masses. All of our linewidths are from the \hi profile when viewing the galaxy edge-on ($i = 90^{\circ}$), and therefore are inclination-corrected. 
   
     Errors on the lowest-mass galaxy are based on percent errors in single-dish data, and this is discussed further in Appendix \ref{errorbar}. We show the variance in our simulations on the right side of the plot for W$_\text{10}$ and W$_\text{20}$ for each mass bin spanning 1 order of magnitude. \hi linewidths vary depending on the viewing angle and mass of the galaxy. We calculate the difference between the linewidth when the galaxy is viewed edge-on versus at a random inclination angle, and then average over all of the galaxies in a given mass bin. Despite the uncertainty on inclination angle for the lower-mass, irregular galaxies, the variance error bars are smallest in this regime as the Gaussian \hi profiles do not vary much with viewing angle. In this regime, the variance is small compared to the estimated observational errors.

     For each velocity method, the information under the legends in Figure \ref{BTFR} 
     shows fits to the equation
    \begin{equation}
        \log{\left( \text{M}_\text{bary} [\text{M}_\odot] \right) } = s \log{\left( V [\text{km s}^{-1}]\right)} + I\text{,}
        \label{btfrEQ}
    \end{equation}
    where $s$ and $I$ are free parameters. The fitted parameters and errors are determined from \verb|numpy.polyfit|. For our vertical scatter, or ordinary least squares error ($\sigma_o$), we use the definition from \citet{2019Lelli} which sums over $N$ galaxies (index $j$):
    \begin{equation}
        \sigma_o = \text{OLS error} = \sqrt{\frac{1}{N}\sum_{j}^{N} \left[ \log(\text{M}_{\text{bary},j}) - s\log{(V_j)} - I \right]^{2}}\text{.}
    \end{equation}
    The fits are shown as dotted colored lines in each panel of Figure \ref{BTFR}.
    
    In summary, a turndown in the bTFR is predicted no matter the method used to find the \textit{halo's} maximum rotation speed, either the mass-enclosed (black points, top left panel) or gravitational midplane potential (black points, top right panel). However, the maximum halo value is not generally recoverable with \hi, whether for spatially resolved methods at the limit $\Sigma_\text{out} = 1 $ M$_\odot$ pc$^{-2}$, or the unresolved \hi linewidths. For resolved rotation curves, using the enclosed mass to estimate the speed results in only slightly higher rotation speeds for lower-mass galaxies as compared to using the midplane potential (at most, $\sim15\%$ difference), but neither shows a turndown when measuring the maximum resolved \hi velocity. For the unresolved \hi profiles, no measurement is able to recover the underlying turndown in the bTFR. These results have major implications for observational studies, as discussed in Section \ref{btfrobs}.  

\subsection{Comparing Velocity Methods and their Radial Extent}
      
    \begin{figure*}
        \centering
        \includegraphics[width=0.9\textwidth]{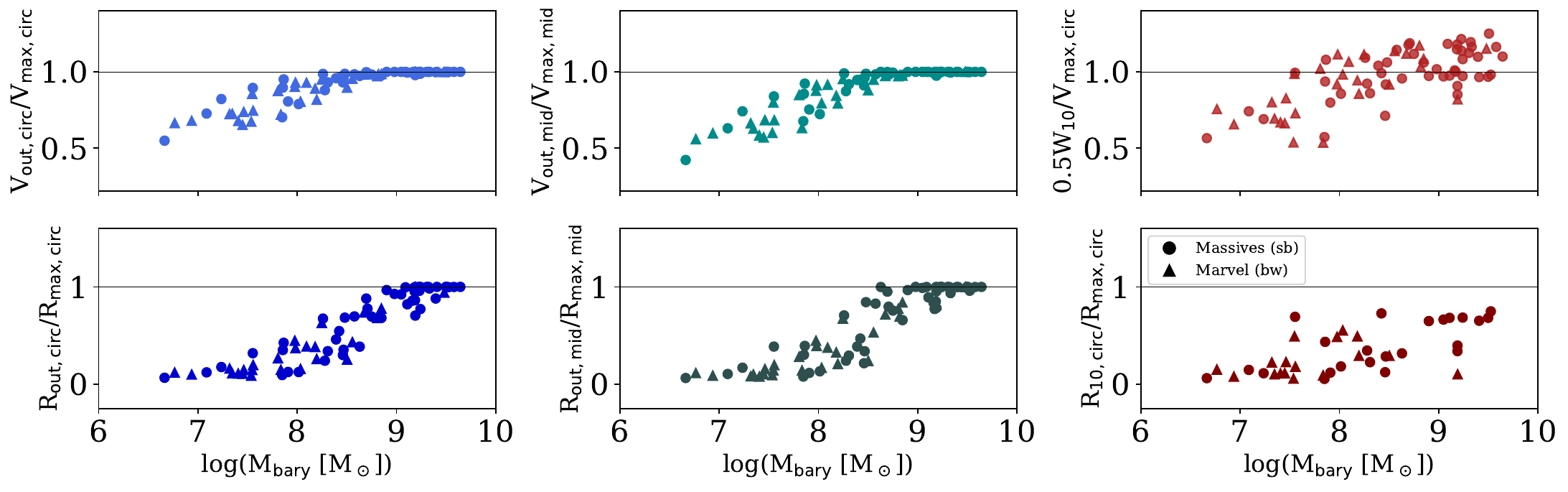}
        \caption{Velocity (top row) and radius (bottom row) ratios to compare each \hi method to the maximum halo values (using the circular velocity curve) as a function of baryonic mass. The left and middle columns show the spatially resolved methods, while the right column shows an unresolved method, W$_\text{10}/2$. When the velocity ratios reach close to 1, this suggests that the \hi size (at the fiducial limit of 1 M$_\odot$ pc$^{-2}$) has reached the flat part of the rotation curve, where the max halo speed occurs. At M$_\text{bary} \lesssim 10^{8.5}$ M$_\odot$, the size of the gas disk decreases with decreasing M$_\text{bary}$ such that the \hi velocities significantly underestimate the halo speed.  W$_\text{10}/2$ exhibits the most scatter in how well it can be used to estimate V$_\text{max}$ or R$_\text{max}$.}
        \label{comparison}
    \end{figure*} 

    In Figure \ref{comparison}, we examine the radial extent probed by each method. We plot ratios of the \hi velocities and radii relative to the halo values as a function of baryonic mass. Both spatially resolved methods (left and middle columns in Figure \ref{comparison}) underpredict the maximum halo speed and corresponding radius by very similar factors at a given baryonic mass. With galaxies at M$_\text{bary} > 10^{8.5}$ M$_{\odot}$, the \hi size at the 1 M$_\odot$ pc$^{-2}$ limit (R$_\text{1}$) extends to the `flat' part of the rotation curve. This is apparent with the velocity ratios being close to 1 above this mass. At lower masses, the \hi size shrinks as galaxy mass decreases, and limits the radial extent of spatially resolved rotation curves. As the \hi shrinks, it no longer extends to the radius where the halo's maximum velocity is reached.  This shrinking of \hi size is the primary reason that the spatially resolved \hi methods under-predict the maximum rotation speed of the lowest-mass galaxies.
    
    In the right column of Figure \ref{comparison}, we compare the max velocity and corresponding radius of W$_\text{10}$/2 to the halo value.  We measure an equivalent radius (R$_\text{10,circ}$) corresponding to the (mass-enclosed) circular velocity curve at a value equal to W$_\text{10}$/2. Compared to V$_\text{out, circ}$ and V$_\text{out, mid}$, W$_\text{10}$ has more vertical scatter in its velocity and radius ratios. Since we plot all \hi linewidths at $i\approx90^\circ$, this vertical scatter is likely not due to inclination, but perhaps from diversity in \hi profile shapes (see Figure \ref{allhi}). We will study the impact of dispersion on \hi linewidths in future work.

    In Appendix \ref{sec:VmaxmidEQ}, we also determine best-fit relationships between the \hi velocities and the maximum rotation speeds, V$_\text{max,mid}$ and V$_\text{max,circ}$, which could potentially be used to extrapolate observational results.

\subsection{bTFR with different \hi sensitivity limits}

    \begin{figure*}
    \includegraphics[width=0.85\textwidth]{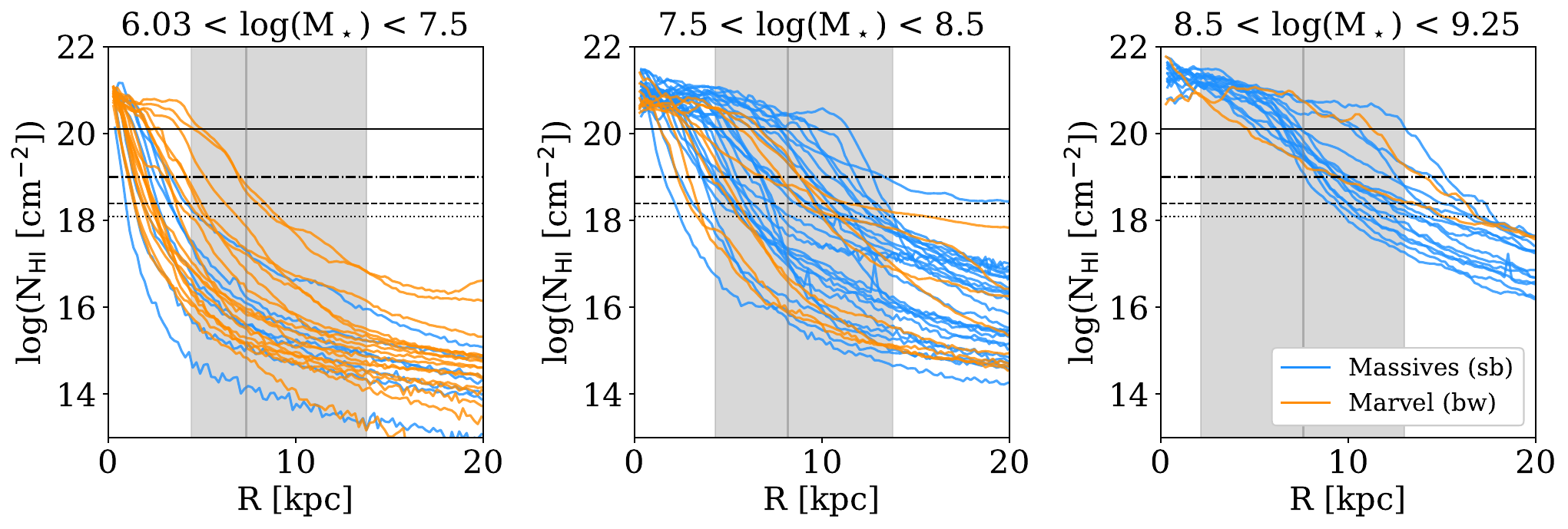}
    \caption{Column density profiles for three stellar mass bins. Each galaxy's column density profile is colored by simulation, where blue is for the Massive Dwarfs suite and orange is for the Marvel suite. The black horizontal lines correspond to four surface density limits that we use to measure the bTFR: 1.0  M$_\odot$ pc$^{-2}$ ($1.25\times10^{20}$ cm$^{-2}$) is the solid line, 0.08 M$_\odot$ pc$^{-2}$ ($1\times10^{19}$ cm$^{-2}$) is the dotted-dashed line, 0.02  M$_\odot$ pc$^{-2}$ ($2.5\times10^{18}$ cm$^{-2}$) is the dashed line, and 0.01  M$_\odot$ pc$^{-2}$ ($1.25\times10^{18}$ cm$^{-2}$) is the dotted line. The shaded grey region shows the range of R$_\text{max,circ}$ values for the galaxies in that mass bin, and the vertical grey line shows the mean value.}
    \label{btfrdens}
\end{figure*}

    We determined that our galaxies with M$_\text{bary} > 10^{8.5}$M$_\odot$ have canonical \hi disk sizes which have reached the flat part of the rotation curve. However, the \hi disk size is sensitive to the surface density limit in observations \citep[e.g.,][]{Wang2025}.  So far we have used an \hi limit of $\Sigma_\text{out} = 1 $ M$_\odot$ pc$^{-2}$, which is the standard definition used in the literature. New and future \hi surveys can push to deeper limits (as discussed in Section \ref{btfrobs}).  In this section, we explore the sensitivities required to trace R$_\text{max,mid}$ and subsequently V$_\text{max,mid}$.

    Figure \ref{btfrdens} shows the \hi column density profiles in three stellar mass bins for our simulated dwarf galaxies. Horizontal lines correspond to the \hi surface density limits: 1.0, 0.08, 0.02, and 0.01M$_\odot$pc$^{-2}$. Our column density profiles plateau with radius, but with different plateaus as a function of mass. More massive galaxies (M$_\star \gtrsim 10^8$M$_\odot$) plateau around $10^{17}$ cm$^{-2}$, and less massive galaxies plateau around $10^{14}$ cm$^{-2}$. These column density profiles are calculated with \textsc{Pynbody} using radial shells. Using the same simulations, \citet{2025Piacitelli} calculated column density profiles with a different method, using sightlines throughout the circumgalactic medium. We have verified that we are consistent out to the 20 kpc shown here. Our simulated galaxies exhibit the full range of \hi column densities found in other simulations ($10^{14} - 10^{21}$ cm$^{-2}$), in which the lower-density gas corresponds to cold accretion from the intergalactic medium \citep{2005Keres,2009Popping, Cook2024}. 

\begin{figure}
    \centering
    \includegraphics[width=0.5\textwidth]{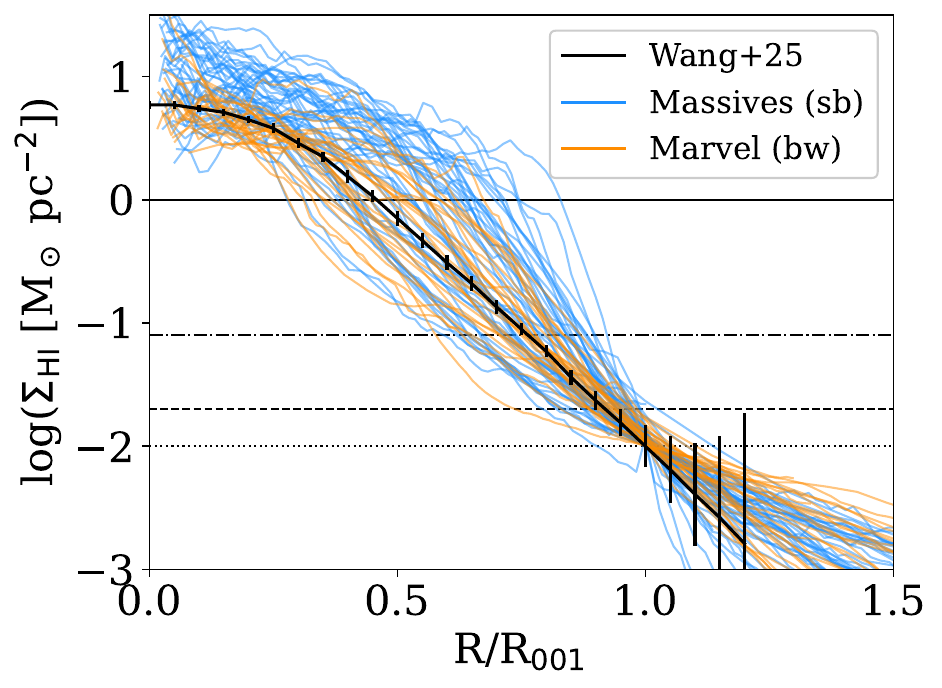}
    \caption{Comparison of normalized \hi surface density profiles from our simulated sample (orange and blue lines) versus the median profile of the FEASTS sample \citep{Wang2025} (black curve). The horizontal lines represent the same column density limits as Figure \ref{btfrdens}. The observed median profile is consistent with the range of our simulated profiles, though constraints at further radii are needed to confirm or rule out the more extended \hi gas in our simulations.}
    \label{comparefeasts}
\end{figure}

    In Figure \ref{comparefeasts}, we compare our surface density profiles to the median surface density profile for the FEASTS sample \citep{Wang2025}. This profile is normalized by R$_\text{001}$, which is the \hi size at a surface density limit of 0.01 M$_\odot$ pc$^{-2}$.   The FEASTS sample targets more massive galaxies, with M$_\text{HI}\gtrsim 10^{8.5}$ M$_\odot$ and M$_\star\gtrsim 10^{8.5}$ M$_\odot$, while most of our galaxies are below this mass range.  Thus, it is not clear if this comparison is valid, but the normalized R/R$_\text{001}$ span two orders of magnitude in galaxy mass and show uniformity, suggesting extrapolation to lower masses might be acceptable.  With this caveat, we find that the observed median profile lies within the range of our simulated surface density profiles.  Galaxies from the Massive Dwarfs simulations appear to have excess surface density of \hi in their inner regions compared to observations.  Discrepancies may be due to the feedback and radiative transfer models used in our simulations.  For example, simulations have a hard time maintaining molecular gas, which wants to immediately form star particles, so must be regulated by feedback.  Our higher \hi surface densities in the central regions of galaxies are likely because some of this gas should be molecular, but feedback turns it to \hi instead. Overall, it is clear that the simulations tend to follow the observational trend in normalized \hi surface density profile at large radii.  Additional observations of surface density profiles out to large radii like  R$_\text{001}$ will help us to calibrate the gas physics.

\begin{figure*}
    \includegraphics[width=0.62\textwidth]{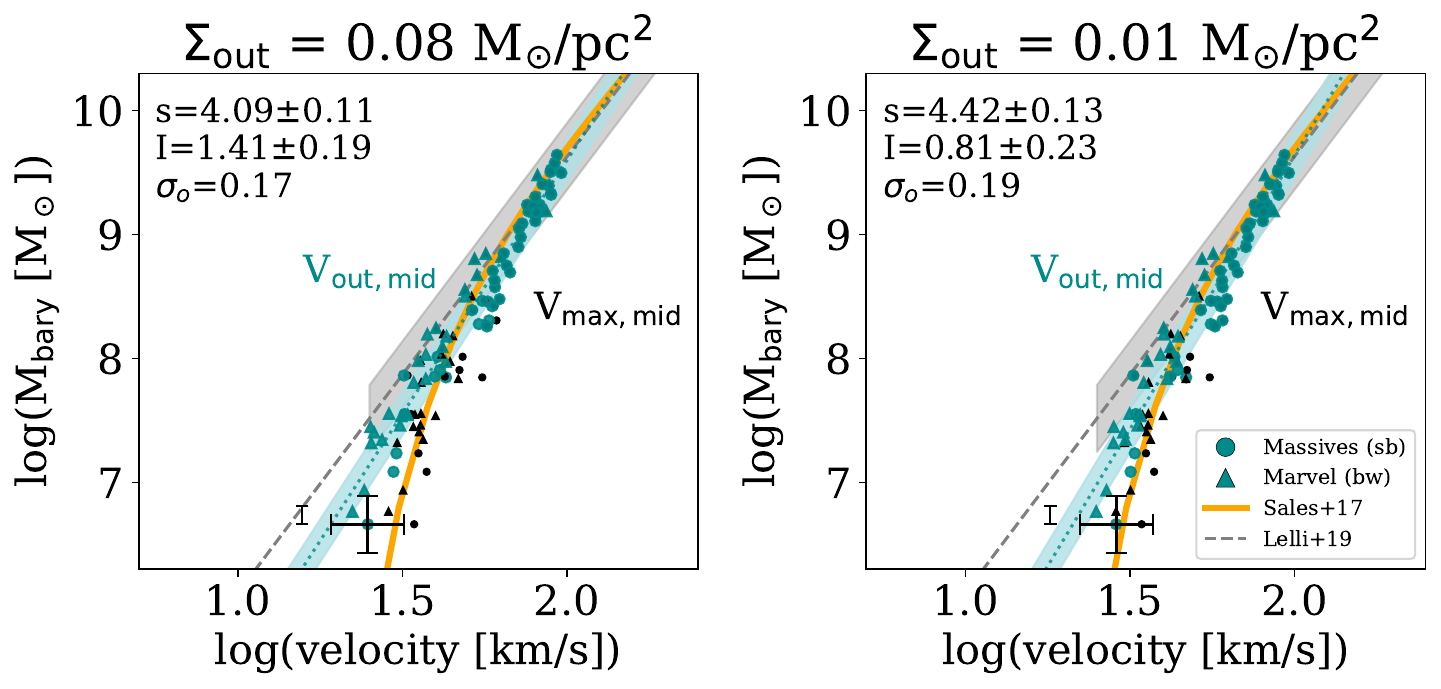}
    \includegraphics[width=0.35\textwidth]{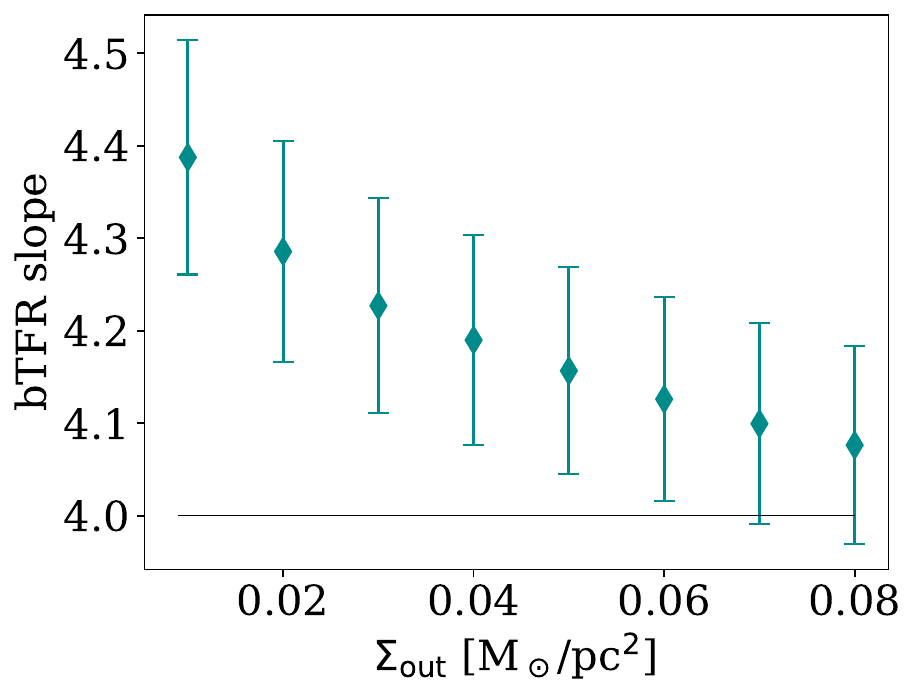}
    \caption{The bTFR using V$_\text{out,mid}$ (dark cyan points) and V$_\text{max,mid}$ (black points) with different \hi sensitivity limits: left panel is 0.08 M$_\odot$ pc$^{-2}$ ($1\times10^{19}$ cm$^{-2}$) and the middle panel is 0.01  M$_\odot$ pc$^{-2}$ ($1.25\times10^{18}$ cm$^{-2}$). The dotted blue line shows a power-law fit to V$_\text{out,mid}$, and the shaded blue region represents the 1-$\sigma$ vertical scatter.  Improving the sensitivity limit to $<0.08$  M$_\odot$ pc$^{-2}$ measures far out enough in radius such that V$_\text{out,mid}$ is within error of V$_\text{max,mid}$, at least for our galaxy sample. Errors on the lowest-mass galaxy are the same as in Figure \ref{BTFR} (top panels). The right panel demonstrates how the bTFR slope is > 4 and steepens as the outer \hi surface density limit improves from 0.08 M$_\odot$ pc$^{-2}$ to 0.01 M$_\odot$ pc$^{-2}$.}
    \label{btfrsens}
\end{figure*}
    
    In Figure \ref{btfrsens}, we compare the bTFR using V$_\text{out,mid}$ at different surface density limits. Improving the sensitivity and reaching a lower \hi column density does indeed begin to close the gap between V$_\text{out,mid}$ and V$_\text{max,mid}$. At 0.08 M$_\odot$ pc$^{-2}$, V$_\text{out,mid}$ recovers V$_\text{max,mid}$ down to 45 km s$^{-1}$. This leads to a change in the bTFR slope as a function of \hi surface density limit, which can be tested by observations. We fit one power-law slope for our sample of galaxies (with V$_\text{max,mid}=28-122$ km s$^{-1}$). For 0.08 M$_\odot$ pc$^{-2}$, the bTFR slope becomes steeper with a value of $4.09\pm0.11$, and steeper still for 0.01 M$_\odot$ pc$^{-2}$ with a value of $4.42 \pm 0.13$. As demonstrated by the lowest limits, a slope $>4.0$ may indicate a turndown in the bTFR. The bTFR slope may be more indicative of our sensitivity limits than alternate models of gravity, as we can measure a slope $\sim4$ within $\Lambda$CDM. Measuring the outermost rotation velocity at lower surface densities allows us to recover the maximum halo speed, though sufficient spatial resolution is also needed. At a surface density limit of 0.01 M$_\odot$ pc$^{-2}$, our sample has \hi sizes ranging from $1.6 - 20$ kpc.

\section{Discussion}
\label{discussion}

\subsection{Comparing bTFR Slopes}
\label{btfrslopes}

For the spatially resolved velocity methods, our bTFR slope using V$_\text{out,circ}$ ($3.78 \pm 0.1$) is slightly steeper than studies in a similar baryonic mass range, while the slope using V$_\text{out,mid}$ ($3.39 \pm 0.09$) is more consistent with observations. \citet{2019Lelli} found a bTFR slope of $3.52 \pm 0.07$, and \citet{2024Siljeg} found that classical dwarfs observed with Apertif followed a similar slope. \citet{2024Deg} used early WALLABY observations and found a slope of $3.3 \pm 0.2$. Our slopes are less steep than the slope from \citet{2009Stark} ($3.94 \pm 0.07$ (random) $\pm 0.08$ (systematic)), and this is likely due to their selection of gas-rich galaxies that reach a flat rotation curve, which has been shown to produce a slope closer to 4 \citep{2016BradfordBTFR}. 

Dwarf galaxies exhibit more non-circular orbits due to baryonic feedback. With more dispersion-supported galaxies in the lower mass range, the discrepancy between assuming the mass-enclosed potential (V$_\text{out,circ}$) versus the midplane gravitational potential (V$_\text{out,circ}$) becomes larger. This may explain the steeper bTFR slope for V$_\text{out,circ}$ ($3.78 \pm 0.1$) versus V$_\text{out,mid}$ ($3.39 \pm 0.09$). Non-circular motions are discussed further in Section \ref{sec:midplane}, and will be explored in a future work. 

For the spatially unresolved methods, our bTFR slope for W$_\text{50}$ ($2.65\pm0.15$) is shallower, while the slopes for W$_\text{20}$ ($3.25\pm0.12$) and W$_\text{10}$ ($3.55\pm0.12$) are closer to observations. Using W$_\text{20}$, \citet{2016BradfordBTFR} measured a slope of $3.24 \pm 0.05$, and \citet{2019Lelli} measured $3.75 \pm 0.08$. We expect W$_\text{10}$ and W$_\text{20}$ to have steeper bTFR slopes than W$_\text{50}$, since W$_\text{50}$ < W$_\text{20}$ < W$_\text{10}$ based on the shape of \hi profiles, which become more Gaussian as mass decreases. This is consistent with previous studies \citep{2016BradfordBTFR,ElBadry2017,2019Lelli,2024Sardone}. Our bTFR slope using W$_\text{20}$ is consistent with \citet{2016BradfordBTFR} and may differ from other studies due to how we treat inclination angle (as discussed in Section \ref{inclination}) and the fact that we include more irregular, lower-mass galaxies in our sample.

\subsection{Non-circular motions and the bTFR}
\label{sec:midplane}

Although our spatially resolved bTFR slopes are generally consistent with observations (Section \ref{btfrslopes}), we consider factors which could affect our rotation curve analysis.

Rotation curves measured in observations and simulations may not fully account for the non-circular motions of dwarf galaxies. Observers typically use software such as $^\mathrm{3D}$\textsc{BAROLO} \citep{2015DiTeodoro} or \textsc{FAT} \citep{2007Jozsa,2015Kamphuis} to model rotation curves from emission spectra. These codes iteratively model the galaxy with tilted three-dimensional rings to match observed kinematic information, and free parameters such as inclination and position angle can be constrained. In our simulations, we measure rotation curves when using the mass enclosed by spherical shells or otherwise calculate the velocity based on the radial component of the gravitational acceleration of the entire galaxy on a few sample points in the midplane. Measuring rotation curves in observations or simulations require some assumptions for smoothness and symmetry, though not all galaxies have well-ordered, thin, gaseous rotating disks.  A body of existing work suggests that deviations from this ideal systematically act to reduce the rotation velocities compared to that measured by enclosed mass \citep[e.g.,][]{Read2016, ElBadry2017, Verbeke2017, Pineda2017, Oman2019, Roper2023, Downing2023, Sands2024}, and that non-circular velocities become increasingly important as mass decreases \citep{Marasco2018}. The bursty star formation in dwarf galaxies can cause fluctuations in the gravitational potential that drive the galaxy out of equilibrium \citep[e.g.,][]{2012Pontzen}, which can impact the rotation curve \citep{Read2016}.  The feedback from these bursts drive non-circular motions such as turbulence or outflows, or the creation of \hi holes that may not be discernible depending on the inclination angle of the observed galaxy \citep{Read2016, Verbeke2017, Sands2024}.  \citet{Marasco2018} shows that triaxial halos can induce bars (radial gas flows) in dwarf galaxies, and the majority of our simulated galaxies with M$_{\star} < 10^8$ M$_{\odot}$ reside in triaxial halos \citep{2025Keith}.  Gas can also experience pressure support that reduces the velocity relative to stars and dark matter, and the pressure support may come from thermal, magnetic, or cosmic ray pressure \citep{Sands2024}.  Thermal pressure support is generally thought to be most relevant in the central regions \citep{Pineda2017}, but effects from pressure support may uniquely influence the \hi size and mass in lower mass galaxies. Disk thickness also plays an increasingly important role as mass decreases \citep{ManceraPina2022} since disks become thicker with decreasing mass \citep[e.g.,][]{SanchezJanseen2010, Roychowdhury2013}, and this can cause overlapping velocities between modeled thin rings.  Most of these effects tend to lower the observed rotation velocity further than what we measure with either of our mass enclosed or midplane potential methods.

This uncertainty is why dwarf galaxies are often selected to be well-behaved before being included in velocity studies. Previous modelers have suggested that a galaxy's disk may be out of equilibrium if the \hi line-of-sight velocity dispersion is $> 20$ km s$^{-1}$ \citep{Verbeke2017}, or not well modeled if the disk height-to-radius ratio is $ > 0.3$ \citep{Sands2024} or inclination is $< 40^\circ$ \citep{Read2016}. However, it has been found that discarding galaxies that are not well-behaved leads to a steeper bTFR slope \citep[e.g.,][]{2009Stark, Downing2023}. As previously mentioned, we have chosen to use all of our galaxies despite their morphology, which may explain discrepancies between our bTFR results and that of observations. In addition to this, it is likely that our bTFR slopes would change if we were able to measure the rotation speeds with non-circular motions fully taken into account. We already see this in the V$_\text{out,mid}$ bTFR result, where measuring the rotation speed with the midplane potential gives a lower velocity than using the enclosed mass, presumably due to increasing importance of non-circular motions, and gives a slope of 3.39 instead of 3.78.

\subsection{Can we measure V$_\text{max}$ in dwarf galaxies with future observations?}
    \label{btfrobs}
    Measuring V$_\text{max}$ for dwarf galaxies with spatially resolved observations will first require a larger sample with sufficiently low rotation speeds. Similar to ALFALFA, the WALLABY, FASHI \citep[FAST all sky \hi survey;][]{2024Zhang}, and CRAFTS \citep[Commensal Radio Astronomy FasT Survey;][]{2021Zhang} surveys will be the next deep, wide-field \hi surveys to discover many more galaxies. WALLABY is expected to cover 14000 deg$^2$ \citep{2012Duffy,2024Murugeshan}, while FASHI and CRAFTS will cover $\sim22000$ deg$^2$ \citep{2021Zhang,2024Zhang}, and each survey is expected to detect $\sim10^5$ galaxies. WALLABY will cover a portion of the southern sky, in the declination range $-60^\circ < \delta< -15^{\circ}$ \citep{2024Murugeshan}, while other surveys will cover portions of the northern sky, like FASHI/CRAFTS with declinations $-14^\circ < \delta < +66^\circ$ \citep{2021Zhang,2024Zhang} and Apertif with $+30^{\circ}<\delta<+60^{\circ}$ \citep{2022Adams}. In addition to observing more galaxies with large sky coverage, MIGHTEE-\hi \citep{2021Maddox} will cover a deeper cosmological volume up to redshift of 0.6. 
    Combining these surveys to identify lower-mass galaxies and conduct follow-up for the kinematics could yield a decent sample of dwarf galaxies to study the bTFR. The WALLABY pilot survey already measured the bTFR for 61 galaxies down to rotation speeds $\sim 35$ km s$^{-1}$ \citep{2024Deg}, but no turndown was detected. However, it is  also worth noting that their sample does not include any galaxies with M$_\text{bary} < 10^8$M$_\odot$, which is the regime where we predict a turndown, and they used an \hi surface density limit of 1 M$_\odot$ pc$^{-2}$.

    Furthermore, to measure V$_\text{max}$ in dwarf galaxies with spatially resolved observations, we need sufficiently low \hi surface density limits. Our results suggest that solely observing galaxies at low rotation speeds is not enough to observe a turndown in the bTFR if we use the 1 M$_\odot$ pc$^{-2}$ surface density limit.  Instead, measuring the rotation of diffuse \hi gas at low densities is crucial. Figure \ref{btfrsens} demonstrates that better sensitivity with a limit $\lesssim 0.08$ M$_\odot$pc$^{-2}$ is needed to possibly measure the bTFR turndown with \hi. Surveys like MHONGOOSE \citep{2024deBlok} and FEASTS \citep{Wang2025} focus on measuring the diffuse \hi gas in the outskirts of galaxies, down to surface densities $\sim0.0039$ M$_\odot$ pc$^{-2}$. MHONGOOSE has a sample of 30 galaxies, and FEASTS is designed to study 118 galaxies\footnote{\url{https://github.com/FEASTS/LVgal/wiki}}, both with M$_\text{HI}\gtrsim10^7$ M$_\odot$. Although MHONGOOSE and FEASTS have a limited sample of dwarf galaxies, V$_\text{max}$ could be successfully measured. Going deeper with MIGHTEE-\hi or surveys with the Square Kilometer Array - Mid \citep{2022Swart} could also increase the sample of galaxies with kinematic information beyond the 1 M$_\odot$ pc$^{-2}$ limit.  If future \hi surveys measure V$_\text{out}$ for more dwarf galaxies at deeper sensitivities and find a steeper bTFR slope ($>4$), this may confirm a turndown. Observing a turndown in the bTFR would provide evidence that indeed, the inefficient galaxy formation of lower-mass galaxies alters the scaling between an \hi disk and dark matter halo, and would suggest that this scaling is not the same in all mass regimes.
     
    Lastly, from Figure \ref{BTFR}, it is not clear that it is possible to measure V$_\text{max}$ using unresolved \hi linewidths, as even W$_\text{10}$ shows little evidence for a turndown.  Certainly the bTFR slope steepens as we move from a measurement at W$_\text{20}$ to W$_\text{10}$ ($3.25 \pm 0.12$ to $3.55 \pm 0.12$). This work focuses on how different velocity definitions impact the bTFR, and therefore we have not performed mock observations or beam smearing effects. Yet even with these `ideal' \hi profiles, with decent spectral resolution (2.6 km s$^{-1}$) and no observational noise, V$_\text{max}$ is not within error of W$_\text{10}$ for our simulated galaxies at M$_\text{bary} \lesssim 10^{8.5}$M$_\odot$. This result is in striking contrast to the general wisdom that unresolved linewidths will be necessary to obtain statistics at low masses. Instead, these results suggest we might have to rely on a handful of well-measured resolved velocities at low galaxy masses in order to probe for evidence of a bTFR turndown.  

\section{Conclusions}
\label{conclusions}

We analyze the bTFR for a sample of 75 simulated dwarf galaxies from the Marvel-ous and Marvelous Massive Dwarfs simulations. These galaxies span a mass range of M$_\star = 10^6-10^9$ M$_\odot$. They are are primarily gas-dominated (Figure \ref{gasdom}) and follow observed scaling relations for \hi gas content and velocity widths (Figures \ref{HIsizemass} and \ref{sizes}).  

Like previous galaxy simulations with CDM \citep[e.g.,][]{2012Ferrero, 2017Sales}, we predict a turndown in the bTFR when the true maximum rotation velocity of the halo is measured (black points, Figure \ref{BTFR}). The bTFR turndown occurs at M$_\text{bary} \lesssim 10^{8.5}$M$_\odot$, or a corresponding rotation speed $\lesssim 50$ km s$^{-1}$. However, we determine the rotation velocity in other ways, including measuring within the outermost \hi radius (at 1 M$_\odot$ pc$^{-2}$) using the mass enclosed in spherical shells (V$_\text{out,circ}$), the gravitational potential in the midplane of the galaxy (V$_\text{out,mid}$), and various widths of the unresolved \hi line profile (W$_\text{50}$, W$_\text{20}$, W$_\text{10}$). We find that none of these more observationally-oriented velocity measurements can recover the underlying turndown in the bTFR.

There are a few physical reasons why spatially resolved velocities might not trace the true V$_\text{max,circ}$ of the halo.  We summarize two key findings:
    \begin{itemize}
        \item Spatially resolved velocities (at the 1 M$_\odot$ pc$^{-2}$ limit) fail to trace V$_\text{max}$ below M$_\text{bary} \sim 10^{8.5}$ M$_\odot$ (Figure \ref{BTFR}) because the size of the \hi disk shrinks as mass declines (see Figure \ref{comparison}). As the disk shrinks, the velocity is measured further in on the rising part of the rotation curve, reducing the outermost measured velocity.

        \item Our \hi column density profiles predict that more gas can be observed past the fiducial limit of 1 M$_\odot$ pc$^{-2}$ (Figure \ref{btfrdens}), though more observations are needed to confirm the extended \hi at radii beyond the 0.01 M$_\odot$pc$^{-2}$ limit (Figure \ref{comparefeasts}). Better sensitivity is necessary to observe the more diffuse gas. We predict that a turndown in the bTFR within $\Lambda$CDM (detectable as a slope $> 4$) could be measured using V$_\text{out,mid}$ if an \hi sensitivity limit of $\lesssim$ 0.08 M$_\odot$ pc$^{-2}$ is achieved (Figure \ref{btfrsens}).  This sensitivity is within reach of measurements in MIGHTEE-HI, MHONGOOSE, and FEASTS, though the number of observed dwarf galaxies that reach this depth may be small overall.
        
    \end{itemize}

It is notable that our W$_\text{20}$ unresolved \hi linewidths are not sensitive to a bTFR turndown, and instead continue the power-law trend from higher mass galaxies.  A steeper slope might be recoverable if W$_\text{10}$ is measured, but this requires incredibly deep integrations with high signal-to-noise \citep{2024Sardone}, and will be difficult to achieve for more dwarf galaxies.  It is generally assumed that unresolved \hi linewidths must be used in order to obtain statistical samples of \hi velocities at the dwarf scale.  However, our results suggest that fewer dwarfs with resolved rotation curves to lower \hi surface densities might be the only way to observationally recover a turndown in the bTFR.

\section*{Acknowledgements}

Resources supporting this work were provided by the NASA High-End Computing (HEC) Program through the NASA Advanced Supercomputing (NAS) Division at Ames Research Center. Some of the simulations were performed using resources made available by the Flatiron Institute. The Flatiron Institute is a division of the Simons Foundation. This work used Stampede2 at the Texas Advanced Computing Center (TACC) through allocation MCA94P018 from the Advanced Cyberinfrastructure Coordination Ecosystem: Services \& Support (ACCESS) program, which is supported by U.S. National Science Foundation grants \#2138259, \#2138286, \#2138307, \#2137603, and \#2138296. A.H.G.P. is supported by NSF Grant No. AST-2008110. J.W. is supported by a grant from NSERC (National Science and Engineering Research Council) Canada.

\section*{Data Availability Statement}
Data from the Marvel-ous and Massive Dwarfs simulations have not been publicly released, but can be provided upon request to the author(s).

\bibliographystyle{mnras}
\bibliography{main}

\begin{thebibliography}{}
\makeatletter
\relax
\def\mn@urlcharsother{\let\do\@makeother \do\$\do\&\do\#\do\^\do\_\do\%\do\~}
\def\mn@doi{\begingroup\mn@urlcharsother \@ifnextchar [ {\mn@doi@} {\mn@doi@[]}}
\def\mn@doi@[#1]#2{\def\@tempa{#1}\ifx\@tempa\@empty \href {http://dx.doi.org/#2} {doi:#2}\else \href {http://dx.doi.org/#2} {#1}\fi \endgroup}
\def\mn@eprint#1#2{\mn@eprint@#1:#2::\@nil}
\def\mn@eprint@arXiv#1{\href {http://arxiv.org/abs/#1} {{\tt arXiv:#1}}}
\def\mn@eprint@dblp#1{\href {http://dblp.uni-trier.de/rec/bibtex/#1.xml} {dblp:#1}}
\def\mn@eprint@#1:#2:#3:#4\@nil{\def\@tempa {#1}\def\@tempb {#2}\def\@tempc {#3}\ifx \@tempc \@empty \let \@tempc \@tempb \let \@tempb \@tempa \fi \ifx \@tempb \@empty \def\@tempb {arXiv}\fi \@ifundefined {mn@eprint@\@tempb}{\@tempb:\@tempc}{\expandafter \expandafter \csname mn@eprint@\@tempb\endcsname \expandafter{\@tempc}}}

\bibitem[\protect\citeauthoryear{{Abel}, {Anninos}, {Zhang}  \& {Norman}}{{Abel} et~al.}{1997}]{1997Abel}
{Abel} T.,  {Anninos} P.,  {Zhang} Y.,   {Norman} M.~L.,  1997, \mn@doi [\na] {10.1016/S1384-1076(97)00010-9}, \href {https://ui.adsabs.harvard.edu/abs/1997NewA....2..181A} {2, 181}

\bibitem[\protect\citeauthoryear{{Adams} et~al.,}{{Adams} et~al.}{2022}]{2022Adams}
{Adams} E.~A.~K.,  et~al., 2022, \mn@doi [\aap] {10.1051/0004-6361/202244007}, \href {https://ui.adsabs.harvard.edu/abs/2022A&A...667A..38A} {667, A38}

\bibitem[\protect\citeauthoryear{{Arnett}}{{Arnett}}{1996}]{1996Arnett}
{Arnett} D.,  1996, {Supernovae and Nucleosynthesis: An Investigation of the History of Matter from the Big Bang to the Present}

\bibitem[\protect\citeauthoryear{{Azartash-Namin} et~al.,}{{Azartash-Namin} et~al.}{2024}]{2024arXiv240106041A}
{Azartash-Namin} B.,  et~al., 2024, \mn@doi [\apj] {10.3847/1538-4357/ad49a5}, \href {https://ui.adsabs.harvard.edu/abs/2024ApJ...970...40A} {970, 40}

\bibitem[\protect\citeauthoryear{{Barnes} \& {Efstathiou}}{{Barnes} \& {Efstathiou}}{1987}]{1987Barnes}
{Barnes} J.,  {Efstathiou} G.,  1987, \mn@doi [\apj] {10.1086/165480}, \href {https://ui.adsabs.harvard.edu/abs/1987ApJ...319..575B} {319, 575}

\bibitem[\protect\citeauthoryear{{Begum}, {Chengalur}, {Karachentsev}, {Sharina}  \& {Kaisin}}{{Begum} et~al.}{2008}]{2008Begum}
{Begum} A.,  {Chengalur} J.~N.,  {Karachentsev} I.~D.,  {Sharina} M.~E.,   {Kaisin} S.~S.,  2008, \mn@doi [\mnras] {10.1111/j.1365-2966.2008.13150.x}, \href {https://ui.adsabs.harvard.edu/abs/2008MNRAS.386.1667B} {386, 1667}

\bibitem[\protect\citeauthoryear{{Behroozi}, {Wechsler}  \& {Conroy}}{{Behroozi} et~al.}{2013}]{2013Behroozi}
{Behroozi} P.~S.,  {Wechsler} R.~H.,   {Conroy} C.,  2013, \mn@doi [\apj] {10.1088/0004-637X/770/1/57}, \href {https://ui.adsabs.harvard.edu/abs/2013ApJ...770...57B} {770, 57}

\bibitem[\protect\citeauthoryear{{Black}}{{Black}}{1981}]{1981Black}
{Black} J.~H.,  1981, \mn@doi [\mnras] {10.1093/mnras/197.3.553}, \href {https://ui.adsabs.harvard.edu/abs/1981MNRAS.197..553B} {197, 553}

\bibitem[\protect\citeauthoryear{{Bosma}}{{Bosma}}{1978}]{1978Bosma}
{Bosma} A.,  1978, PhD thesis, University of Groningen, Netherlands

\bibitem[\protect\citeauthoryear{{Bradford}, {Geha}  \& {Blanton}}{{Bradford} et~al.}{2015}]{2015ApJ...809..146B}
{Bradford} J.~D.,  {Geha} M.~C.,   {Blanton} M.~R.,  2015, \mn@doi [\apj] {10.1088/0004-637X/809/2/146}, \href {https://ui.adsabs.harvard.edu/abs/2015ApJ...809..146B} {809, 146}

\bibitem[\protect\citeauthoryear{{Bradford}, {Geha}  \& {van den Bosch}}{{Bradford} et~al.}{2016}]{2016BradfordBTFR}
{Bradford} J.~D.,  {Geha} M.~C.,   {van den Bosch} F.~C.,  2016, \mn@doi [\apj] {10.3847/0004-637X/832/1/11}, \href {https://ui.adsabs.harvard.edu/abs/2016ApJ...832...11B} {832, 11}

\bibitem[\protect\citeauthoryear{{Broeils} \& {Rhee}}{{Broeils} \& {Rhee}}{1997}]{1997BandRhee}
{Broeils} A.~H.,  {Rhee} M.~H.,  1997, \aap, \href {https://ui.adsabs.harvard.edu/abs/1997A&A...324..877B} {324, 877}

\bibitem[\protect\citeauthoryear{{Brook} \& {Shankar}}{{Brook} \& {Shankar}}{2016}]{2016Brook}
{Brook} C.~B.,  {Shankar} F.,  2016, \mn@doi [\mnras] {10.1093/mnras/stv2550}, \href {https://ui.adsabs.harvard.edu/abs/2016MNRAS.455.3841B} {455, 3841}

\bibitem[\protect\citeauthoryear{{Brook}, {Santos-Santos}  \& {Stinson}}{{Brook} et~al.}{2016}]{2016Brook_diffv}
{Brook} C.~B.,  {Santos-Santos} I.,   {Stinson} G.,  2016, \mn@doi [\mnras] {10.1093/mnras/stw650}, \href {https://ui.adsabs.harvard.edu/abs/2016MNRAS.459..638B} {459, 638}

\bibitem[\protect\citeauthoryear{{Brooks}, {Papastergis}, {Christensen}, {Governato}, {Stilp}, {Quinn}  \& {Wadsley}}{{Brooks} et~al.}{2017}]{2017Brooks}
{Brooks} A.~M.,  {Papastergis} E.,  {Christensen} C.~R.,  {Governato} F.,  {Stilp} A.,  {Quinn} T.~R.,   {Wadsley} J.,  2017, \mn@doi [\apj] {10.3847/1538-4357/aa9576}, \href {https://ui.adsabs.harvard.edu/abs/2017ApJ...850...97B} {850, 97}

\bibitem[\protect\citeauthoryear{{Catinella} et~al.,}{{Catinella} et~al.}{2010}]{2010Catinella}
{Catinella} B.,  et~al., 2010, \mn@doi [\mnras] {10.1111/j.1365-2966.2009.16180.x}, \href {https://ui.adsabs.harvard.edu/abs/2010MNRAS.403..683C} {403, 683}

\bibitem[\protect\citeauthoryear{{Cen}}{{Cen}}{1992}]{1992Cen}
{Cen} R.,  1992, \mn@doi [\apjs] {10.1086/191630}, \href {https://ui.adsabs.harvard.edu/abs/1992ApJS...78..341C} {78, 341}

\bibitem[\protect\citeauthoryear{{Christensen}, {Quinn}, {Governato}, {Stilp}, {Shen}  \& {Wadsley}}{{Christensen} et~al.}{2012}]{2012MNRAS.425.3058C}
{Christensen} C.,  {Quinn} T.,  {Governato} F.,  {Stilp} A.,  {Shen} S.,   {Wadsley} J.,  2012, \mn@doi [\mnras] {10.1111/j.1365-2966.2012.21628.x}, \href {https://ui.adsabs.harvard.edu/abs/2012MNRAS.425.3058C} {425, 3058}

\bibitem[\protect\citeauthoryear{{Christensen}, {Governato}, {Quinn}, {Brooks}, {Shen}, {McCleary}, {Fisher}  \& {Wadsley}}{{Christensen} et~al.}{2014}]{Christensen2014}
{Christensen} C.~R.,  {Governato} F.,  {Quinn} T.,  {Brooks} A.~M.,  {Shen} S.,  {McCleary} J.,  {Fisher} D.~B.,   {Wadsley} J.,  2014, \mn@doi [\mnras] {10.1093/mnras/stu399}, \href {https://ui.adsabs.harvard.edu/abs/2014MNRAS.440.2843C} {440, 2843}

\bibitem[\protect\citeauthoryear{{Cook}, {van de Voort}, {Pakmor}  \& {Grand}}{{Cook} et~al.}{2024}]{Cook2024}
{Cook} A. W.~S.,  {van de Voort} F.,  {Pakmor} R.,   {Grand} R. J.~J.,  2024, \mn@doi [arXiv e-prints] {10.48550/arXiv.2409.05578}, \href {https://ui.adsabs.harvard.edu/abs/2024arXiv240905578C} {p. arXiv:2409.05578}

\bibitem[\protect\citeauthoryear{{Cowie} \& {McKee}}{{Cowie} \& {McKee}}{1977}]{1977Cowie}
{Cowie} L.~L.,  {McKee} C.~F.,  1977, \mn@doi [\apj] {10.1086/154911}, \href {https://ui.adsabs.harvard.edu/abs/1977ApJ...211..135C} {211, 135}

\bibitem[\protect\citeauthoryear{{Deg} et~al.,}{{Deg} et~al.}{2024}]{2024Deg}
{Deg} N.,  et~al., 2024, \mn@doi [\apj] {10.3847/1538-4357/ad84ba}, \href {https://ui.adsabs.harvard.edu/abs/2024ApJ...976..159D} {976, 159}

\bibitem[\protect\citeauthoryear{{Di Cintio}, {Brook}, {Macci{\`o}}, {Stinson}, {Knebe}, {Dutton}  \& {Wadsley}}{{Di Cintio} et~al.}{2014}]{2014DiCintio}
{Di Cintio} A.,  {Brook} C.~B.,  {Macci{\`o}} A.~V.,  {Stinson} G.~S.,  {Knebe} A.,  {Dutton} A.~A.,   {Wadsley} J.,  2014, \mn@doi [\mnras] {10.1093/mnras/stt1891}, \href {https://ui.adsabs.harvard.edu/abs/2014MNRAS.437..415D} {437, 415}

\bibitem[\protect\citeauthoryear{{Di Teodoro} \& {Fraternali}}{{Di Teodoro} \& {Fraternali}}{2015}]{2015DiTeodoro}
{Di Teodoro} E.~M.,  {Fraternali} F.,  2015, \mn@doi [\mnras] {10.1093/mnras/stv1213}, \href {https://ui.adsabs.harvard.edu/abs/2015MNRAS.451.3021D} {451, 3021}

\bibitem[\protect\citeauthoryear{{Downing} \& {Oman}}{{Downing} \& {Oman}}{2023}]{Downing2023}
{Downing} E.~R.,  {Oman} K.~A.,  2023, \mn@doi [\mnras] {10.1093/mnras/stad868}, \href {https://ui.adsabs.harvard.edu/abs/2023MNRAS.522.3318D} {522, 3318}

\bibitem[\protect\citeauthoryear{{Duffy}, {Meyer}, {Staveley-Smith}, {Bernyk}, {Croton}, {Koribalski}, {Gerstmann}  \& {Westerlund}}{{Duffy} et~al.}{2012}]{2012Duffy}
{Duffy} A.~R.,  {Meyer} M.~J.,  {Staveley-Smith} L.,  {Bernyk} M.,  {Croton} D.~J.,  {Koribalski} B.~S.,  {Gerstmann} D.,   {Westerlund} S.,  2012, \mn@doi [\mnras] {10.1111/j.1365-2966.2012.21987.x}, \href {https://ui.adsabs.harvard.edu/abs/2012MNRAS.426.3385D} {426, 3385}

\bibitem[\protect\citeauthoryear{{Dutton}}{{Dutton}}{2012}]{2012Dutton}
{Dutton} A.~A.,  2012, \mn@doi [\mnras] {10.1111/j.1365-2966.2012.21469.x}, \href {https://ui.adsabs.harvard.edu/abs/2012MNRAS.424.3123D} {424, 3123}

\bibitem[\protect\citeauthoryear{{Dutton}, {Obreja}  \& {Macci{\`o}}}{{Dutton} et~al.}{2019}]{2019Dutton}
{Dutton} A.~A.,  {Obreja} A.,   {Macci{\`o}} A.~V.,  2019, \mn@doi [\mnras] {10.1093/mnras/sty3064}, \href {https://ui.adsabs.harvard.edu/abs/2019MNRAS.482.5606D} {482, 5606}

\bibitem[\protect\citeauthoryear{{El-Badry}, {Wetzel}, {Geha}, {Quataert}, {Hopkins}, {Kere{\v{s}}}, {Chan}  \& {Faucher-Gigu{\`e}re}}{{El-Badry} et~al.}{2017}]{ElBadry2017}
{El-Badry} K.,  {Wetzel} A.~R.,  {Geha} M.,  {Quataert} E.,  {Hopkins} P.~F.,  {Kere{\v{s}}} D.,  {Chan} T.~K.,   {Faucher-Gigu{\`e}re} C.-A.,  2017, \mn@doi [\apj] {10.3847/1538-4357/835/2/193}, \href {https://ui.adsabs.harvard.edu/abs/2017ApJ...835..193E} {835, 193}

\bibitem[\protect\citeauthoryear{{El-Badry} et~al.,}{{El-Badry} et~al.}{2018}]{ElBadry2018}
{El-Badry} K.,  et~al., 2018, \mn@doi [\mnras] {10.1093/mnras/sty730}, \href {https://ui.adsabs.harvard.edu/abs/2018MNRAS.477.1536E} {477, 1536}

\bibitem[\protect\citeauthoryear{{Ferrero}, {Abadi}, {Navarro}, {Sales}  \& {Gurovich}}{{Ferrero} et~al.}{2012}]{2012Ferrero}
{Ferrero} I.,  {Abadi} M.~G.,  {Navarro} J.~F.,  {Sales} L.~V.,   {Gurovich} S.,  2012, \mn@doi [\mnras] {10.1111/j.1365-2966.2012.21623.x}, \href {https://ui.adsabs.harvard.edu/abs/2012MNRAS.425.2817F} {425, 2817}

\bibitem[\protect\citeauthoryear{{Geha}, {Blanton}, {Masjedi}  \& {West}}{{Geha} et~al.}{2006}]{2006Geha}
{Geha} M.,  {Blanton} M.~R.,  {Masjedi} M.,   {West} A.~A.,  2006, \mn@doi [\apj] {10.1086/508604}, \href {https://ui.adsabs.harvard.edu/abs/2006ApJ...653..240G} {653, 240}

\bibitem[\protect\citeauthoryear{{Gill}, {Knebe}  \& {Gibson}}{{Gill} et~al.}{2004}]{2004MNRAS.351..399G}
{Gill} S. P.~D.,  {Knebe} A.,   {Gibson} B.~K.,  2004, \mn@doi [\mnras] {10.1111/j.1365-2966.2004.07786.x}, \href {https://ui.adsabs.harvard.edu/abs/2004MNRAS.351..399G} {351, 399}

\bibitem[\protect\citeauthoryear{{Girardi} et~al.,}{{Girardi} et~al.}{2010}]{2010Girardi}
{Girardi} L.,  et~al., 2010, \mn@doi [\apj] {10.1088/0004-637X/724/2/1030}, \href {https://ui.adsabs.harvard.edu/abs/2010ApJ...724.1030G} {724, 1030}

\bibitem[\protect\citeauthoryear{{Glowacki}, {Elson}  \& {Dav{\'e}}}{{Glowacki} et~al.}{2020}]{2020Glowacki}
{Glowacki} M.,  {Elson} E.,   {Dav{\'e}} R.,  2020, \mn@doi [\mnras] {10.1093/mnras/staa2616}, \href {https://ui.adsabs.harvard.edu/abs/2020MNRAS.498.3687G} {498, 3687}

\bibitem[\protect\citeauthoryear{{Gurovich}, {Freeman}, {Jerjen}, {Staveley-Smith}  \& {Puerari}}{{Gurovich} et~al.}{2010}]{2010Gurovich}
{Gurovich} S.,  {Freeman} K.,  {Jerjen} H.,  {Staveley-Smith} L.,   {Puerari} I.,  2010, \mn@doi [\aj] {10.1088/0004-6256/140/3/663}, \href {https://ui.adsabs.harvard.edu/abs/2010AJ....140..663G} {140, 663}

\bibitem[\protect\citeauthoryear{{Haardt} \& {Madau}}{{Haardt} \& {Madau}}{2012}]{2012HaardyMadau}
{Haardt} F.,  {Madau} P.,  2012, \mn@doi [\apj] {10.1088/0004-637X/746/2/125}, \href {https://ui.adsabs.harvard.edu/abs/2012ApJ...746..125H} {746, 125}

\bibitem[\protect\citeauthoryear{{Haynes} et~al.,}{{Haynes} et~al.}{2011}]{2011Haynes}
{Haynes} M.~P.,  et~al., 2011, \mn@doi [\aj] {10.1088/0004-6256/142/5/170}, \href {https://ui.adsabs.harvard.edu/abs/2011AJ....142..170H} {142, 170}

\bibitem[\protect\citeauthoryear{{Haynes} et~al.,}{{Haynes} et~al.}{2018}]{2018Haynes}
{Haynes} M.~P.,  et~al., 2018, \mn@doi [\apj] {10.3847/1538-4357/aac956}, \href {https://ui.adsabs.harvard.edu/abs/2018ApJ...861...49H} {861, 49}

\bibitem[\protect\citeauthoryear{{Hunter} et~al.,}{{Hunter} et~al.}{2012}]{2012Hunter}
{Hunter} D.~A.,  et~al., 2012, \mn@doi [\aj] {10.1088/0004-6256/144/5/134}, \href {https://ui.adsabs.harvard.edu/abs/2012AJ....144..134H} {144, 134}

\bibitem[\protect\citeauthoryear{{Hunter}, {Elmegreen}  \& {Madden}}{{Hunter} et~al.}{2024}]{2024Hunter}
{Hunter} D.~A.,  {Elmegreen} B.~G.,   {Madden} S.~C.,  2024, \mn@doi [\araa] {10.1146/annurev-astro-052722-104109}, \href {https://ui.adsabs.harvard.edu/abs/2024ARA&A..62..113H} {62, 113}

\bibitem[\protect\citeauthoryear{{J{\'o}zsa}, {Kenn}, {Klein}  \& {Oosterloo}}{{J{\'o}zsa} et~al.}{2007}]{2007Jozsa}
{J{\'o}zsa} G.~I.~G.,  {Kenn} F.,  {Klein} U.,   {Oosterloo} T.~A.,  2007, \mn@doi [\aap] {10.1051/0004-6361:20066164}, \href {https://ui.adsabs.harvard.edu/abs/2007A&A...468..731J} {468, 731}

\bibitem[\protect\citeauthoryear{Kale \& Krishnan}{Kale \& Krishnan}{1993}]{KaleKrishnan}
Kale L.,  Krishnan S.,  1993, \mn@doi [ACM SIGPLAN Notices] {10.1145/167962.165874}, 28, 91

\bibitem[\protect\citeauthoryear{{Kamphuis}, {J{\'o}zsa}, {Oh}, {Spekkens}, {Urbancic}, {Serra}, {Koribalski}  \& {Dettmar}}{{Kamphuis} et~al.}{2015}]{2015Kamphuis}
{Kamphuis} P.,  {J{\'o}zsa} G.~I.~G.,  {Oh} S. .~H.,  {Spekkens} K.,  {Urbancic} N.,  {Serra} P.,  {Koribalski} B.~S.,   {Dettmar} R.~J.,  2015, \mn@doi [\mnras] {10.1093/mnras/stv1480}, \href {https://ui.adsabs.harvard.edu/abs/2015MNRAS.452.3139K} {452, 3139}

\bibitem[\protect\citeauthoryear{{Keith} et~al.,}{{Keith} et~al.}{2025}]{2025Keith}
{Keith} B.,  et~al., 2025, \mn@doi [\apj] {10.3847/1538-4357/add40d}, \href {https://ui.adsabs.harvard.edu/abs/2025ApJ...986..138K} {986, 138}

\bibitem[\protect\citeauthoryear{{Keller}, {Wadsley}, {Benincasa}  \& {Couchman}}{{Keller} et~al.}{2014}]{2014MNRAS.442.3013K}
{Keller} B.~W.,  {Wadsley} J.,  {Benincasa} S.~M.,   {Couchman} H.~M.~P.,  2014, \mn@doi [\mnras] {10.1093/mnras/stu1058}, \href {https://ui.adsabs.harvard.edu/abs/2014MNRAS.442.3013K} {442, 3013}

\bibitem[\protect\citeauthoryear{{Kennicutt}}{{Kennicutt}}{1998}]{1998Kennicutt}
{Kennicutt} Jr. R.~C.,  1998, \mn@doi [\araa] {10.1146/annurev.astro.36.1.189}, \href {https://ui.adsabs.harvard.edu/abs/1998ARA&A..36..189K} {36, 189}

\bibitem[\protect\citeauthoryear{{Kere{\v{s}}}, {Katz}, {Weinberg}  \& {Dav{\'e}}}{{Kere{\v{s}}} et~al.}{2005}]{2005Keres}
{Kere{\v{s}}} D.,  {Katz} N.,  {Weinberg} D.~H.,   {Dav{\'e}} R.,  2005, \mn@doi [\mnras] {10.1111/j.1365-2966.2005.09451.x}, \href {https://ui.adsabs.harvard.edu/abs/2005MNRAS.363....2K} {363, 2}

\bibitem[\protect\citeauthoryear{{Klypin}, {Karachentsev}, {Makarov}  \& {Nasonova}}{{Klypin} et~al.}{2015}]{2015Klypin}
{Klypin} A.,  {Karachentsev} I.,  {Makarov} D.,   {Nasonova} O.,  2015, \mn@doi [\mnras] {10.1093/mnras/stv2040}, \href {https://ui.adsabs.harvard.edu/abs/2015MNRAS.454.1798K} {454, 1798}

\bibitem[\protect\citeauthoryear{{Knollmann} \& {Knebe}}{{Knollmann} \& {Knebe}}{2009}]{2009ApJS..182..608K}
{Knollmann} S.~R.,  {Knebe} A.,  2009, \mn@doi [\apjs] {10.1088/0067-0049/182/2/608}, \href {https://ui.adsabs.harvard.edu/abs/2009ApJS..182..608K} {182, 608}

\bibitem[\protect\citeauthoryear{{Koribalski} et~al.,}{{Koribalski} et~al.}{2020}]{2020Koribalski}
{Koribalski} B.~S.,  et~al., 2020, \mn@doi [\apss] {10.1007/s10509-020-03831-4}, \href {https://ui.adsabs.harvard.edu/abs/2020Ap&SS.365..118K} {365, 118}

\bibitem[\protect\citeauthoryear{{Kroupa}}{{Kroupa}}{2001}]{2001Kroupa}
{Kroupa} P.,  2001, \mn@doi [\mnras] {10.1046/j.1365-8711.2001.04022.x}, \href {https://ui.adsabs.harvard.edu/abs/2001MNRAS.322..231K} {322, 231}

\bibitem[\protect\citeauthoryear{{Krumm} \& {Burstein}}{{Krumm} \& {Burstein}}{1984}]{1984Krumm}
{Krumm} N.,  {Burstein} D.,  1984, \mn@doi [\aj] {10.1086/113630}, \href {https://ui.adsabs.harvard.edu/abs/1984AJ.....89.1319K} {89, 1319}

\bibitem[\protect\citeauthoryear{{Lelli}, {McGaugh}  \& {Schombert}}{{Lelli} et~al.}{2016a}]{2016Lelli}
{Lelli} F.,  {McGaugh} S.~S.,   {Schombert} J.~M.,  2016a, \mn@doi [\aj] {10.3847/0004-6256/152/6/157}, \href {https://ui.adsabs.harvard.edu/abs/2016AJ....152..157L} {152, 157}

\bibitem[\protect\citeauthoryear{{Lelli}, {McGaugh}  \& {Schombert}}{{Lelli} et~al.}{2016b}]{2016LelliBTFR}
{Lelli} F.,  {McGaugh} S.~S.,   {Schombert} J.~M.,  2016b, \mn@doi [\apjl] {10.3847/2041-8205/816/1/L14}, \href {https://ui.adsabs.harvard.edu/abs/2016ApJ...816L..14L} {816, L14}

\bibitem[\protect\citeauthoryear{{Lelli}, {McGaugh}, {Schombert}, {Desmond}  \& {Katz}}{{Lelli} et~al.}{2019}]{2019Lelli}
{Lelli} F.,  {McGaugh} S.~S.,  {Schombert} J.~M.,  {Desmond} H.,   {Katz} H.,  2019, \mn@doi [\mnras] {10.1093/mnras/stz205}, \href {https://ui.adsabs.harvard.edu/abs/2019MNRAS.484.3267L} {484, 3267}

\bibitem[\protect\citeauthoryear{{Macci{\`o}}, {Udrescu}, {Dutton}, {Obreja}, {Wang}, {Stinson}  \& {Kang}}{{Macci{\`o}} et~al.}{2016}]{2016Maccio}
{Macci{\`o}} A.~V.,  {Udrescu} S.~M.,  {Dutton} A.~A.,  {Obreja} A.,  {Wang} L.,  {Stinson} G.~R.,   {Kang} X.,  2016, \mn@doi [\mnras] {10.1093/mnrasl/slw147}, \href {https://ui.adsabs.harvard.edu/abs/2016MNRAS.463L..69M} {463, L69}

\bibitem[\protect\citeauthoryear{{Maddox} et~al.,}{{Maddox} et~al.}{2021}]{2021Maddox}
{Maddox} N.,  et~al., 2021, \mn@doi [\aap] {10.1051/0004-6361/202039655}, \href {https://ui.adsabs.harvard.edu/abs/2021A&A...646A..35M} {646, A35}

\bibitem[\protect\citeauthoryear{{Mancera Pi{\~n}a}, {Fraternali}, {Oosterloo}, {Adams}, {di Teodoro}, {Bacchini}  \& {Iorio}}{{Mancera Pi{\~n}a} et~al.}{2022}]{ManceraPina2022}
{Mancera Pi{\~n}a} P.~E.,  {Fraternali} F.,  {Oosterloo} T.,  {Adams} E. A.~K.,  {di Teodoro} E.,  {Bacchini} C.,   {Iorio} G.,  2022, \mn@doi [\mnras] {10.1093/mnras/stac1508}, \href {https://ui.adsabs.harvard.edu/abs/2022MNRAS.514.3329M} {514, 3329}

\bibitem[\protect\citeauthoryear{{Marasco}, {Oman}, {Navarro}, {Frenk}  \& {Oosterloo}}{{Marasco} et~al.}{2018}]{Marasco2018}
{Marasco} A.,  {Oman} K.~A.,  {Navarro} J.~F.,  {Frenk} C.~S.,   {Oosterloo} T.,  2018, \mn@doi [\mnras] {10.1093/mnras/sty354}, \href {https://ui.adsabs.harvard.edu/abs/2018MNRAS.476.2168M} {476, 2168}

\bibitem[\protect\citeauthoryear{{Marigo}, {Girardi}, {Bressan}, {Groenewegen}, {Silva}  \& {Granato}}{{Marigo} et~al.}{2008}]{2008Marigo}
{Marigo} P.,  {Girardi} L.,  {Bressan} A.,  {Groenewegen} M.~A.~T.,  {Silva} L.,   {Granato} G.~L.,  2008, \mn@doi [\aap] {10.1051/0004-6361:20078467}, \href {https://ui.adsabs.harvard.edu/abs/2008A&A...482..883M} {482, 883}

\bibitem[\protect\citeauthoryear{{McGaugh}}{{McGaugh}}{2012}]{2012McGaugh}
{McGaugh} S.~S.,  2012, \mn@doi [\aj] {10.1088/0004-6256/143/2/40}, \href {https://ui.adsabs.harvard.edu/abs/2012AJ....143...40M} {143, 40}

\bibitem[\protect\citeauthoryear{{McGaugh}, {Schombert}, {Bothun}  \& {de Blok}}{{McGaugh} et~al.}{2000}]{2000McGaugh}
{McGaugh} S.~S.,  {Schombert} J.~M.,  {Bothun} G.~D.,   {de Blok} W.~J.~G.,  2000, \mn@doi [\apjl] {10.1086/312628}, \href {https://ui.adsabs.harvard.edu/abs/2000ApJ...533L..99M} {533, L99}

\bibitem[\protect\citeauthoryear{{McQuinn} et~al.,}{{McQuinn} et~al.}{2022}]{2022McQuinn}
{McQuinn} K. B.~W.,  et~al., 2022, \mn@doi [\apj] {10.3847/1538-4357/ac9285}, \href {https://ui.adsabs.harvard.edu/abs/2022ApJ...940....8M} {940, 8}

\bibitem[\protect\citeauthoryear{{Menon}, {Wesolowski}, {Zheng}, {Jetley}, {Kale}, {Quinn}  \& {Governato}}{{Menon} et~al.}{2015}]{2015ComAC...2....1M}
{Menon} H.,  {Wesolowski} L.,  {Zheng} G.,  {Jetley} P.,  {Kale} L.,  {Quinn} T.,   {Governato} F.,  2015, \mn@doi [Computational Astrophysics and Cosmology] {10.1186/s40668-015-0007-9}, \href {https://ui.adsabs.harvard.edu/abs/2015ComAC...2....1M} {2, 1}

\bibitem[\protect\citeauthoryear{{Milgrom}}{{Milgrom}}{1983}]{1983Milgrom}
{Milgrom} M.,  1983, \mn@doi [\apj] {10.1086/161131}, \href {https://ui.adsabs.harvard.edu/abs/1983ApJ...270..371M} {270, 371}

\bibitem[\protect\citeauthoryear{{Mina}, {Shen}, {Keller}, {Mayer}, {Madau}  \& {Wadsley}}{{Mina} et~al.}{2021}]{2021Mina}
{Mina} M.,  {Shen} S.,  {Keller} B.~W.,  {Mayer} L.,  {Madau} P.,   {Wadsley} J.,  2021, \mn@doi [\aap] {10.1051/0004-6361/202039420}, \href {https://ui.adsabs.harvard.edu/abs/2021A&A...655A..22M} {655, A22}

\bibitem[\protect\citeauthoryear{{Moster}, {Naab}  \& {White}}{{Moster} et~al.}{2013}]{2013Moster}
{Moster} B.~P.,  {Naab} T.,   {White} S. D.~M.,  2013, \mn@doi [\mnras] {10.1093/mnras/sts261}, \href {https://ui.adsabs.harvard.edu/abs/2013MNRAS.428.3121M} {428, 3121}

\bibitem[\protect\citeauthoryear{{Munshi} et~al.,}{{Munshi} et~al.}{2013}]{2013Munshi}
{Munshi} F.,  et~al., 2013, \mn@doi [\apj] {10.1088/0004-637X/766/1/56}, \href {https://ui.adsabs.harvard.edu/abs/2013ApJ...766...56M} {766, 56}

\bibitem[\protect\citeauthoryear{{Munshi}, {Brooks}, {Applebaum}, {Christensen}, {Quinn}  \& {Sligh}}{{Munshi} et~al.}{2021}]{2021ApJ...923...35M}
{Munshi} F.,  {Brooks} A.~M.,  {Applebaum} E.,  {Christensen} C.~R.,  {Quinn} T.,   {Sligh} S.,  2021, \mn@doi [\apj] {10.3847/1538-4357/ac0db6}, \href {https://ui.adsabs.harvard.edu/abs/2021ApJ...923...35M} {923, 35}

\bibitem[\protect\citeauthoryear{{Murugeshan} et~al.,}{{Murugeshan} et~al.}{2024}]{2024Murugeshan}
{Murugeshan} C.,  et~al., 2024, \mn@doi [\pasa] {10.1017/pasa.2024.91}, \href {https://ui.adsabs.harvard.edu/abs/2024PASA...41...88M} {41, e088}

\bibitem[\protect\citeauthoryear{{Oh} et~al.,}{{Oh} et~al.}{2015}]{2015Oh}
{Oh} S.-H.,  et~al., 2015, \mn@doi [\aj] {10.1088/0004-6256/149/6/180}, \href {https://ui.adsabs.harvard.edu/abs/2015AJ....149..180O} {149, 180}

\bibitem[\protect\citeauthoryear{{Oman}, {Navarro}, {Sales}, {Fattahi}, {Frenk}, {Sawala}, {Schaller}  \& {White}}{{Oman} et~al.}{2016}]{Oman2016}
{Oman} K.~A.,  {Navarro} J.~F.,  {Sales} L.~V.,  {Fattahi} A.,  {Frenk} C.~S.,  {Sawala} T.,  {Schaller} M.,   {White} S. D.~M.,  2016, \mn@doi [\mnras] {10.1093/mnras/stw1251}, \href {https://ui.adsabs.harvard.edu/abs/2016MNRAS.460.3610O} {460, 3610}

\bibitem[\protect\citeauthoryear{{Oman}, {Marasco}, {Navarro}, {Frenk}, {Schaye}  \& {Ben{\'\i}tez-Llambay}}{{Oman} et~al.}{2019}]{Oman2019}
{Oman} K.~A.,  {Marasco} A.,  {Navarro} J.~F.,  {Frenk} C.~S.,  {Schaye} J.,   {Ben{\'\i}tez-Llambay} A.,  2019, \mn@doi [\mnras] {10.1093/mnras/sty2687}, \href {https://ui.adsabs.harvard.edu/abs/2019MNRAS.482..821O} {482, 821}

\bibitem[\protect\citeauthoryear{{Ostriker} \& {McKee}}{{Ostriker} \& {McKee}}{1988}]{1988blastwave}
{Ostriker} J.~P.,  {McKee} C.~F.,  1988, \mn@doi [Reviews of Modern Physics] {10.1103/RevModPhys.60.1}, \href {https://ui.adsabs.harvard.edu/abs/1988RvMP...60....1O} {60, 1}

\bibitem[\protect\citeauthoryear{{Papastergis} \& {Shankar}}{{Papastergis} \& {Shankar}}{2016}]{2016PapastergisTBTF}
{Papastergis} E.,  {Shankar} F.,  2016, \mn@doi [\aap] {10.1051/0004-6361/201527854}, \href {https://ui.adsabs.harvard.edu/abs/2016A&A...591A..58P} {591, A58}

\bibitem[\protect\citeauthoryear{{Papastergis}, {Adams}  \& {van der Hulst}}{{Papastergis} et~al.}{2016}]{2016Papastergis}
{Papastergis} E.,  {Adams} E.~A.~K.,   {van der Hulst} J.~M.,  2016, \mn@doi [\aap] {10.1051/0004-6361/201628410}, \href {https://ui.adsabs.harvard.edu/abs/2016A&A...593A..39P} {593, A39}

\bibitem[\protect\citeauthoryear{{Peebles}}{{Peebles}}{1969}]{1969Peebles}
{Peebles} P.~J.~E.,  1969, \mn@doi [\apj] {10.1086/149876}, \href {https://ui.adsabs.harvard.edu/abs/1969ApJ...155..393P} {155, 393}

\bibitem[\protect\citeauthoryear{{Piacitelli} et~al.,}{{Piacitelli} et~al.}{2025}]{2025Piacitelli}
{Piacitelli} D.~R.,  et~al., 2025, \mn@doi [arXiv e-prints] {10.48550/arXiv.2505.08861}, \href {https://ui.adsabs.harvard.edu/abs/2025arXiv250508861P} {p. arXiv:2505.08861}

\bibitem[\protect\citeauthoryear{{Pineda}, {Hayward}, {Springel}  \& {Mendes de Oliveira}}{{Pineda} et~al.}{2017}]{Pineda2017}
{Pineda} J. C.~B.,  {Hayward} C.~C.,  {Springel} V.,   {Mendes de Oliveira} C.,  2017, \mn@doi [\mnras] {10.1093/mnras/stw3004}, \href {https://ui.adsabs.harvard.edu/abs/2017MNRAS.466...63P} {466, 63}

\bibitem[\protect\citeauthoryear{{Planck Collaboration} et~al.,}{{Planck Collaboration} et~al.}{2016}]{2016A&A...594A..13P}
{Planck Collaboration} et~al., 2016, \mn@doi [\aap] {10.1051/0004-6361/201525830}, \href {https://ui.adsabs.harvard.edu/abs/2016A&A...594A..13P} {594, A13}

\bibitem[\protect\citeauthoryear{{Planck Collaboration} et~al.,}{{Planck Collaboration} et~al.}{2020}]{2020Planck}
{Planck Collaboration} et~al., 2020, \mn@doi [\aap] {10.1051/0004-6361/201833910}, \href {https://ui.adsabs.harvard.edu/abs/2020A&A...641A...6P} {641, A6}

\bibitem[\protect\citeauthoryear{{Pontzen} \& {Governato}}{{Pontzen} \& {Governato}}{2012}]{2012Pontzen}
{Pontzen} A.,  {Governato} F.,  2012, \mn@doi [\mnras] {10.1111/j.1365-2966.2012.20571.x}, \href {https://ui.adsabs.harvard.edu/abs/2012MNRAS.421.3464P} {421, 3464}

\bibitem[\protect\citeauthoryear{{Pontzen}, {Ro{\v{s}}kar}, {Stinson}  \& {Woods}}{{Pontzen} et~al.}{2013}]{2013ascl.soft05002P}
{Pontzen} A.,  {Ro{\v{s}}kar} R.,  {Stinson} G.,   {Woods} R.,  2013, {pynbody: N-Body/SPH analysis for python}, Astrophysics Source Code Library, record ascl:1305.002

\bibitem[\protect\citeauthoryear{{Popping}, {Dav{\'e}}, {Braun}  \& {Oppenheimer}}{{Popping} et~al.}{2009}]{2009Popping}
{Popping} A.,  {Dav{\'e}} R.,  {Braun} R.,   {Oppenheimer} B.~D.,  2009, \mn@doi [\aap] {10.1051/0004-6361/200911811}, \href {https://ui.adsabs.harvard.edu/abs/2009A&A...504...15P} {504, 15}

\bibitem[\protect\citeauthoryear{{Power}, {Navarro}, {Jenkins}, {Frenk}, {White}, {Springel}, {Stadel}  \& {Quinn}}{{Power} et~al.}{2003}]{2003Power}
{Power} C.,  {Navarro} J.~F.,  {Jenkins} A.,  {Frenk} C.~S.,  {White} S.~D.~M.,  {Springel} V.,  {Stadel} J.,   {Quinn} T.,  2003, \mn@doi [\mnras] {10.1046/j.1365-8711.2003.05925.x}, \href {https://ui.adsabs.harvard.edu/abs/2003MNRAS.338...14P} {338, 14}

\bibitem[\protect\citeauthoryear{{Read}, {Iorio}, {Agertz}  \& {Fraternali}}{{Read} et~al.}{2016}]{Read2016}
{Read} J.~I.,  {Iorio} G.,  {Agertz} O.,   {Fraternali} F.,  2016, \mn@doi [\mnras] {10.1093/mnras/stw1876}, \href {https://ui.adsabs.harvard.edu/abs/2016MNRAS.462.3628R} {462, 3628}

\bibitem[\protect\citeauthoryear{{Read}, {Iorio}, {Agertz}  \& {Fraternali}}{{Read} et~al.}{2017}]{2017Read}
{Read} J.~I.,  {Iorio} G.,  {Agertz} O.,   {Fraternali} F.,  2017, \mn@doi [\mnras] {10.1093/mnras/stx147}, \href {https://ui.adsabs.harvard.edu/abs/2017MNRAS.467.2019R} {467, 2019}

\bibitem[\protect\citeauthoryear{{Roper}, {Oman}, {Frenk}, {Ben{\'\i}tez-Llambay}, {Navarro}  \& {Santos-Santos}}{{Roper} et~al.}{2023}]{Roper2023}
{Roper} F.~A.,  {Oman} K.~A.,  {Frenk} C.~S.,  {Ben{\'\i}tez-Llambay} A.,  {Navarro} J.~F.,   {Santos-Santos} I. M.~E.,  2023, \mn@doi [\mnras] {10.1093/mnras/stad549}, \href {https://ui.adsabs.harvard.edu/abs/2023MNRAS.521.1316R} {521, 1316}

\bibitem[\protect\citeauthoryear{{Roychowdhury}, {Chengalur}, {Karachentsev}  \& {Kaisina}}{{Roychowdhury} et~al.}{2013}]{Roychowdhury2013}
{Roychowdhury} S.,  {Chengalur} J.~N.,  {Karachentsev} I.~D.,   {Kaisina} E.~I.,  2013, \mn@doi [\mnras] {10.1093/mnrasl/slt123}, \href {https://ui.adsabs.harvard.edu/abs/2013MNRAS.436L.104R} {436, L104}

\bibitem[\protect\citeauthoryear{{Rubin}, {Ford}  \& {Thonnard}}{{Rubin} et~al.}{1980}]{1980Rubin}
{Rubin} V.~C.,  {Ford} Jr. W.~K.,   {Thonnard} N.,  1980, \mn@doi [Astrophysical Journal] {10.1086/158003}, \href {https://ui.adsabs.harvard.edu/abs/1980ApJ...238..471R} {238, 471}

\bibitem[\protect\citeauthoryear{{Sales} et~al.,}{{Sales} et~al.}{2017}]{2017Sales}
{Sales} L.~V.,  et~al., 2017, \mn@doi [\mnras] {10.1093/mnras/stw2461}, \href {https://ui.adsabs.harvard.edu/abs/2017MNRAS.464.2419S} {464, 2419}

\bibitem[\protect\citeauthoryear{{S{\'a}nchez-Janssen}, {M{\'e}ndez-Abreu}  \& {Aguerri}}{{S{\'a}nchez-Janssen} et~al.}{2010}]{SanchezJanseen2010}
{S{\'a}nchez-Janssen} R.,  {M{\'e}ndez-Abreu} J.,   {Aguerri} J.~A.~L.,  2010, \mn@doi [\mnras] {10.1111/j.1745-3933.2010.00883.x}, \href {https://ui.adsabs.harvard.edu/abs/2010MNRAS.406L..65S} {406, L65}

\bibitem[\protect\citeauthoryear{{Sands} et~al.,}{{Sands} et~al.}{2024}]{Sands2024}
{Sands} I.~S.,  et~al., 2024, \mn@doi [arXiv e-prints] {10.48550/arXiv.2404.16247}, \href {https://ui.adsabs.harvard.edu/abs/2024arXiv240416247S} {p. arXiv:2404.16247}

\bibitem[\protect\citeauthoryear{{Sardone}, {Peter}, {Brooks}  \& {Kaczmarek}}{{Sardone} et~al.}{2024}]{2024Sardone}
{Sardone} A.,  {Peter} A. H.~G.,  {Brooks} A.~M.,   {Kaczmarek} J.,  2024, \mn@doi [\apj] {10.3847/1538-4357/ad250f}, \href {https://ui.adsabs.harvard.edu/abs/2024ApJ...964..135S} {964, 135}

\bibitem[\protect\citeauthoryear{{Schmidt}}{{Schmidt}}{1959}]{1959Schmidt}
{Schmidt} M.,  1959, \mn@doi [\apj] {10.1086/146614}, \href {https://ui.adsabs.harvard.edu/abs/1959ApJ...129..243S} {129, 243}

\bibitem[\protect\citeauthoryear{{Shen}, {Wadsley}  \& {Stinson}}{{Shen} et~al.}{2010}]{2010MNRAS.407.1581S}
{Shen} S.,  {Wadsley} J.,   {Stinson} G.,  2010, \mn@doi [\mnras] {10.1111/j.1365-2966.2010.17047.x}, \href {https://ui.adsabs.harvard.edu/abs/2010MNRAS.407.1581S} {407, 1581}

\bibitem[\protect\citeauthoryear{{Spekkens}, {Lewis}  \& {Deg}}{{Spekkens} et~al.}{2020}]{2020Spekkens}
{Spekkens} K.,  {Lewis} C.,   {Deg} N.,  2020, in American Astronomical Society Meeting Abstracts \#236. p. 109.05

\bibitem[\protect\citeauthoryear{{Spergel} et~al.,}{{Spergel} et~al.}{2007}]{2007ApJS..170..377S}
{Spergel} D.~N.,  et~al., 2007, \mn@doi [\apjs] {10.1086/513700}, \href {https://ui.adsabs.harvard.edu/abs/2007ApJS..170..377S} {170, 377}

\bibitem[\protect\citeauthoryear{{Stark}, {McGaugh}  \& {Swaters}}{{Stark} et~al.}{2009}]{2009Stark}
{Stark} D.~V.,  {McGaugh} S.~S.,   {Swaters} R.~A.,  2009, \mn@doi [\aj] {10.1088/0004-6256/138/2/392}, \href {https://ui.adsabs.harvard.edu/abs/2009AJ....138..392S} {138, 392}

\bibitem[\protect\citeauthoryear{{Stevens}, {Diemer}, {Lagos}, {Nelson}, {Obreschkow}, {Wang}  \& {Marinacci}}{{Stevens} et~al.}{2019}]{2019Stevens}
{Stevens} A. R.~H.,  {Diemer} B.,  {Lagos} C. d.~P.,  {Nelson} D.,  {Obreschkow} D.,  {Wang} J.,   {Marinacci} F.,  2019, \mn@doi [\mnras] {10.1093/mnras/stz2513}, \href {https://ui.adsabs.harvard.edu/abs/2019MNRAS.490...96S} {490, 96}

\bibitem[\protect\citeauthoryear{{Stilp}, {Dalcanton}, {Warren}, {Skillman}, {Ott}  \& {Koribalski}}{{Stilp} et~al.}{2013a}]{2013StilpB}
{Stilp} A.~M.,  {Dalcanton} J.~J.,  {Warren} S.~R.,  {Skillman} E.,  {Ott} J.,   {Koribalski} B.,  2013a, \mn@doi [\apj] {10.1088/0004-637X/765/2/136}, \href {https://ui.adsabs.harvard.edu/abs/2013ApJ...765..136S} {765, 136}

\bibitem[\protect\citeauthoryear{{Stilp}, {Dalcanton}, {Skillman}, {Warren}, {Ott}  \& {Koribalski}}{{Stilp} et~al.}{2013b}]{2013StilpA}
{Stilp} A.~M.,  {Dalcanton} J.~J.,  {Skillman} E.,  {Warren} S.~R.,  {Ott} J.,   {Koribalski} B.,  2013b, \mn@doi [\apj] {10.1088/0004-637X/773/2/88}, \href {https://ui.adsabs.harvard.edu/abs/2013ApJ...773...88S} {773, 88}

\bibitem[\protect\citeauthoryear{{Stinson}, {Seth}, {Katz}, {Wadsley}, {Governato}  \& {Quinn}}{{Stinson} et~al.}{2006}]{2006MNRAS.373.1074S}
{Stinson} G.,  {Seth} A.,  {Katz} N.,  {Wadsley} J.,  {Governato} F.,   {Quinn} T.,  2006, \mn@doi [\mnras] {10.1111/j.1365-2966.2006.11097.x}, \href {https://ui.adsabs.harvard.edu/abs/2006MNRAS.373.1074S} {373, 1074}

\bibitem[\protect\citeauthoryear{{Swart}, {Dewdney}  \& {Cremonini}}{{Swart} et~al.}{2022}]{2022Swart}
{Swart} G.~P.,  {Dewdney} P.~E.,   {Cremonini} A.,  2022, \mn@doi [Journal of Astronomical Telescopes, Instruments, and Systems] {10.1117/1.JATIS.8.1.011021}, \href {https://ui.adsabs.harvard.edu/abs/2022JATIS...8a1021S} {8, 011021}

\bibitem[\protect\citeauthoryear{{Tamburro}, {Rix}, {Leroy}, {Mac Low}, {Walter}, {Kennicutt}, {Brinks}  \& {de Blok}}{{Tamburro} et~al.}{2009}]{2009Tamburro}
{Tamburro} D.,  {Rix} H.~W.,  {Leroy} A.~K.,  {Mac Low} M.~M.,  {Walter} F.,  {Kennicutt} R.~C.,  {Brinks} E.,   {de Blok} W.~J.~G.,  2009, \mn@doi [\aj] {10.1088/0004-6256/137/5/4424}, \href {https://ui.adsabs.harvard.edu/abs/2009AJ....137.4424T} {137, 4424}

\bibitem[\protect\citeauthoryear{{Tremmel}, {Karcher}, {Governato}, {Volonteri}, {Quinn}, {Pontzen}, {Anderson}  \& {Bellovary}}{{Tremmel} et~al.}{2017}]{2017MNRAS.470.1121T}
{Tremmel} M.,  {Karcher} M.,  {Governato} F.,  {Volonteri} M.,  {Quinn} T.~R.,  {Pontzen} A.,  {Anderson} L.,   {Bellovary} J.,  2017, \mn@doi [\mnras] {10.1093/mnras/stx1160}, \href {https://ui.adsabs.harvard.edu/abs/2017MNRAS.470.1121T} {470, 1121}

\bibitem[\protect\citeauthoryear{{Trujillo-Gomez}, {Schneider}, {Papastergis}, {Reed}  \& {Lake}}{{Trujillo-Gomez} et~al.}{2018}]{2018trujillogomez}
{Trujillo-Gomez} S.,  {Schneider} A.,  {Papastergis} E.,  {Reed} D.~S.,   {Lake} G.,  2018, \mn@doi [\mnras] {10.1093/mnras/sty146}, \href {https://ui.adsabs.harvard.edu/abs/2018MNRAS.475.4825T} {475, 4825}

\bibitem[\protect\citeauthoryear{{Tully} \& {Fisher}}{{Tully} \& {Fisher}}{1977}]{1977TullyFisher}
{Tully} R.~B.,  {Fisher} J.~R.,  1977, \aap, \href {https://ui.adsabs.harvard.edu/abs/1977A&A....54..661T} {54, 661}

\bibitem[\protect\citeauthoryear{{Verbeke}, {Papastergis}, {Ponomareva}, {Rathi}  \& {De Rijcke}}{{Verbeke} et~al.}{2017}]{Verbeke2017}
{Verbeke} R.,  {Papastergis} E.,  {Ponomareva} A.~A.,  {Rathi} S.,   {De Rijcke} S.,  2017, \mn@doi [\aap] {10.1051/0004-6361/201730758}, \href {https://ui.adsabs.harvard.edu/abs/2017A&A...607A..13V} {607, A13}

\bibitem[\protect\citeauthoryear{{Verheijen}}{{Verheijen}}{2001}]{2001Verheijen}
{Verheijen} M. A.~W.,  2001, \mn@doi [\apj] {10.1086/323887}, \href {https://ui.adsabs.harvard.edu/abs/2001ApJ...563..694V} {563, 694}

\bibitem[\protect\citeauthoryear{{Verner} \& {Ferland}}{{Verner} \& {Ferland}}{1996}]{1996Verner}
{Verner} D.~A.,  {Ferland} G.~J.,  1996, \mn@doi [\apjs] {10.1086/192284}, \href {https://ui.adsabs.harvard.edu/abs/1996ApJS..103..467V} {103, 467}

\bibitem[\protect\citeauthoryear{{Wadsley}, {Stadel}  \& {Quinn}}{{Wadsley} et~al.}{2004}]{2004NewA....9..137W}
{Wadsley} J.~W.,  {Stadel} J.,   {Quinn} T.,  2004, \mn@doi [\na] {10.1016/j.newast.2003.08.004}, \href {https://ui.adsabs.harvard.edu/abs/2004NewA....9..137W} {9, 137}

\bibitem[\protect\citeauthoryear{{Wadsley}, {Keller}  \& {Quinn}}{{Wadsley} et~al.}{2017}]{2017MNRAS.471.2357W}
{Wadsley} J.~W.,  {Keller} B.~W.,   {Quinn} T.~R.,  2017, \mn@doi [\mnras] {10.1093/mnras/stx1643}, \href {https://ui.adsabs.harvard.edu/abs/2017MNRAS.471.2357W} {471, 2357}

\bibitem[\protect\citeauthoryear{{Wang}, {Koribalski}, {Serra}, {van der Hulst}, {Roychowdhury}, {Kamphuis}  \& {Chengalur}}{{Wang} et~al.}{2016}]{2016Wang}
{Wang} J.,  {Koribalski} B.~S.,  {Serra} P.,  {van der Hulst} T.,  {Roychowdhury} S.,  {Kamphuis} P.,   {Chengalur} J.~N.,  2016, \mn@doi [\mnras] {10.1093/mnras/stw1099}, \href {https://ui.adsabs.harvard.edu/abs/2016MNRAS.460.2143W} {460, 2143}

\bibitem[\protect\citeauthoryear{{Wang} et~al.,}{{Wang} et~al.}{2025}]{Wang2025}
{Wang} J.,  et~al., 2025, \mn@doi [\apj] {10.3847/1538-4357/ada95a}, \href {https://ui.adsabs.harvard.edu/abs/2025ApJ...980...25W} {980, 25}

\bibitem[\protect\citeauthoryear{{Wechsler} \& {Tinker}}{{Wechsler} \& {Tinker}}{2018}]{Wechsler2018}
{Wechsler} R.~H.,  {Tinker} J.~L.,  2018, \mn@doi [\araa] {10.1146/annurev-astro-081817-051756}, \href {https://ui.adsabs.harvard.edu/abs/2018ARA&A..56..435W} {56, 435}

\bibitem[\protect\citeauthoryear{{Yaryura}, {Helmi}, {Abadi}  \& {Starkenburg}}{{Yaryura} et~al.}{2016}]{2016Yaryura}
{Yaryura} C.~Y.,  {Helmi} A.,  {Abadi} M.~G.,   {Starkenburg} E.,  2016, \mn@doi [\mnras] {10.1093/mnras/stw139}, \href {https://ui.adsabs.harvard.edu/abs/2016MNRAS.457.2415Y} {457, 2415}

\bibitem[\protect\citeauthoryear{{Zhang} et~al.,}{{Zhang} et~al.}{2021}]{2021Zhang}
{Zhang} K.,  et~al., 2021, \mn@doi [\mnras] {10.1093/mnras/staa3275}, \href {https://ui.adsabs.harvard.edu/abs/2021MNRAS.500.1741Z} {500, 1741}

\bibitem[\protect\citeauthoryear{{Zhang} et~al.,}{{Zhang} et~al.}{2024}]{2024Zhang}
{Zhang} C.-P.,  et~al., 2024, \mn@doi [Science China Physics, Mechanics, and Astronomy] {10.1007/s11433-023-2219-7}, \href {https://ui.adsabs.harvard.edu/abs/2024SCPMA..6719511Z} {67, 219511}

\bibitem[\protect\citeauthoryear{{de Blok} et~al.,}{{de Blok} et~al.}{2024}]{2024deBlok}
{de Blok} W.~J.~G.,  et~al., 2024, \mn@doi [\aap] {10.1051/0004-6361/202348297}, \href {https://ui.adsabs.harvard.edu/abs/2024A&A...688A.109D} {688, A109}

\bibitem[\protect\citeauthoryear{{{\v{S}}iljeg} et~al.,}{{{\v{S}}iljeg} et~al.}{2024}]{2024Siljeg}
{{\v{S}}iljeg} B.,  et~al., 2024, \mn@doi [\aap] {10.1051/0004-6361/202449923}, \href {https://ui.adsabs.harvard.edu/abs/2024A&A...692A.217S} {692, A217}

\bibitem[\protect\citeauthoryear{{van Cappellen} et~al.,}{{van Cappellen} et~al.}{2022}]{2022vanCapellen}
{van Cappellen} W.~A.,  et~al., 2022, \mn@doi [\aap] {10.1051/0004-6361/202141739}, \href {https://ui.adsabs.harvard.edu/abs/2022A&A...658A.146V} {658, A146}

\makeatother
\end{thebibliography}

\appendix

\section{\hi Profiles for the Full Sample}
\label{fullsample}
\begin{figure*}
    \centering
    \includegraphics[width=0.55\linewidth]{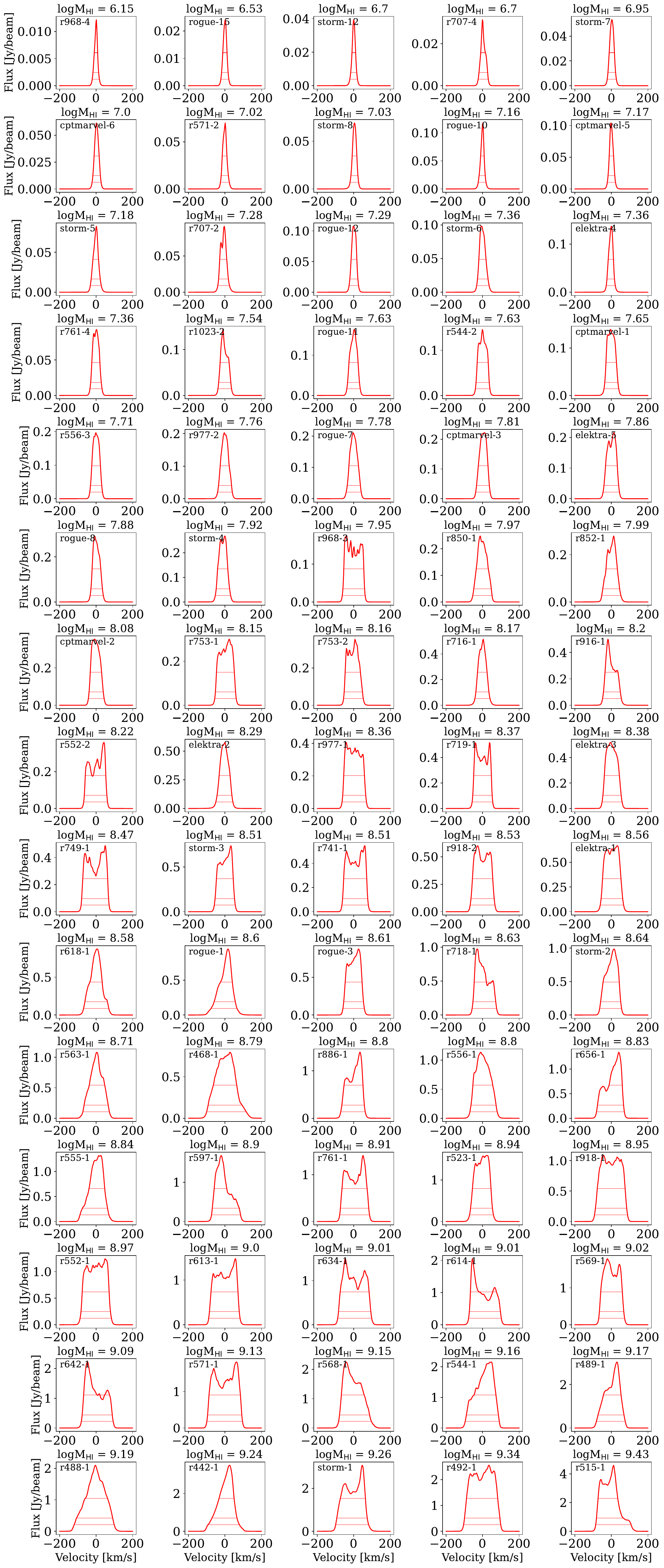}
    \caption{\hi emission profiles for our full sample of 75 simulated dwarf galaxies from the Marvel and Massive Dwarfs suites, ordered by total \hi mass. Each horizontal line corresponds to the linewidths at 50\%, 20\%, and 10\% of the peak flux.}
    \label{allhi}
\end{figure*}

The full sample of 75 dwarf galaxies and their \hi emission profiles are shown in Figure \ref{allhi}. The galaxies are ordered by their total \hi mass and span a range of $\log($M$_\text{\hi}[$M$_\odot])=6.15-9.43$. Galaxies with lower masses tend to have more Gaussian \hi profiles, while higher mass galaxies exhibit the double-horn feature with high-velocity gas rotating in the outermost regions of the disk. Despite the criteria for these galaxies to be isolated (no galaxy of equal or greater mass within the virial radius) and in the field, some of the \hi profiles are asymmetric, which will be explored in future work. 

\section{Estimate of Observational Errors}\label{errorbar}
    The bTFR compares a galaxy's baryonic mass (M$_\text{bary}=$M$_\star$+1.4M$_\text{HI}$) and rotation velocity. Here, we elaborate on our estimate of observational errors for each axis. We do not perform mock observations for our simulated dwarf galaxies. However, we attempt to plot our bTFR trends with reasonable errors to determine the significance of differences between \hi and halo velocities. 

    Our baryonic mass error must account for errors in stellar mass and \hi mass. For stellar mass, we use the mean uncertainty from \citet{2015ApJ...809..146B} of 0.164 dex. This value is chosen since their sample covers a similar mass range with M$_\star=10^7-10^{9.5}$ M$_\star$. For \hi mass errors, we adopt the error function from \citet{2018Haynes}, which propagates distance ($D$) and flux ($S_{21}$) percent uncertainties:
    \begin{equation}
        \sigma_{\log(M_{HI})} = \frac{\sqrt{\left(\frac{\sigma_{S_{21}}}{S_{21}} \right)^2 + \left(  \frac{2\sigma_D}{D} \right)^2 + 0.1^2}}{\ln 10}\text{.}
    \end{equation}
    For this error estimate, we use a signal-to-noise ratio of 10 since this is the threshold used by \citet{2024Sardone} to ensure robust measurements of W$_\text{20}$ and W$_\text{10}$. We adopt a distance uncertainty of $17\%$, which is the average for the SPARC (Spitzer Photometry and Accurate Rotation Curves) galaxies \citep{2016Lelli}. Based on this distance uncertainty, we find an \hi mass uncertainty of 0.161 dex. Since stellar masses and \hi masses are typically measured independently through different methods, we assume Gaussian statistics and take the log-space stellar and \hi mass errors in quadrature to get 0.23 dex as the baryonic mass error. Gaussian statistics are commonly adopted for error propagation in bTFR studies \citep[e.g.,][]{2009Stark, 2016LelliBTFR, 2016Papastergis, 2022McQuinn}.

    For the total error in velocity, we consider the uncertainty from turbulent motion (dispersion) and the uncertainty in measuring the rotation speed. The asymmetric drift correction from gas turbulence is usually approximated with a value of $\sigma_\text{asd} = 8 \pm 2$ km s$^{-1}$ \citep{2013StilpA}. By taking the logarithm of the upper limit as a fraction of the fiducial value, we get $\log\left( 1.25\sigma_\text{asd} / \sigma_\text{asd}\right)= 0.1$ dex (25\% error) as the dispersion uncertainty. For the uncertainty of spatially resolved velocities, we use the average error of $13\%$ from \citet{2022McQuinn}, in which their sample had lower-mass galaxies with M$_\text{bary} < 10^8$ M$_\odot$. \citet{2022McQuinn} similarly assumed Gaussian statistics and added the errors for dispersion and rotation speed in quadrature (Equation 2 of their work), and we find 0.11 dex (29\%) for the velocity error (Figure \ref{BTFR}, top panels). For the uncertainty of \hi linewidths, we use 9\%, which is the average from \citet{2016BradfordBTFR}. Adding the errors in quadrature gives a total velocity uncertainty of 0.10 dex (27\% error), as shown in the bottom row of Figure \ref{BTFR}.  

\section{Relations between Velocity Methods}\label{sec:VmaxmidEQ}

We determine best-fit relationships between the \hi velocities and the maximum rotation speeds, V$_\text{max,mid}$ and V$_\text{max,circ}$. Table \ref{VmaxcircEQ} summarizes the best-fit linear relations between the \hi velocity methods to the halos speeds V$_\text{max,circ}$ (which uses the mass-enclosed gravitational potential) and V$_\text{max,mid}$ (which uses the midplane gravitational potential). In this table, we also include the best-fit linear relation between maximum halo speeds V$_\text{max,mid}$ and V$_\text{max,circ}$.

\begin{table}
    \centering
    \begin{tabular}{|c|c|c|c|}
        \hline
         y & m & x & b \\
         \hline
         V$_\text{out,circ}$ & $1.17 \pm 0.02$ & V$_\text{max,circ}$ & $-13.27 \pm 1.17$ \\
         V$_\text{out,mid}$ & $1.26 \pm 0.03$ & V$_\text{max,circ}$ & $-18.48 \pm 1.58$ \\
         W$_\text{10}/2$ & $1.24 \pm 0.05$ & V$_\text{max,circ}$ & $-14.4 \pm 3.19$ \\
         W$_\text{20}/2$ & $1.15 \pm 0.06$ & V$_\text{max,circ}$ & $-14.8 \pm 3.50$ \\
         W$_\text{50}/2$ & $0.89 \pm 0.08$ & V$_\text{max,circ}$ & $-11.07 \pm 4.76$ \\
         \hline
         \hline
        V$_\text{out,circ}$ & $1.11 \pm 0.02$ & V$_\text{max,mid}$ & $-11.41 \pm 1.13$ \\
         V$_\text{out,mid}$ & $1.20 \pm 0.02$ & V$_\text{max,mid}$ & $-16.63 \pm 1.4$ \\
         W$_\text{10}/2$ & $1.19 \pm 0.05$ & V$_\text{max,mid}$ & $-12.84 \pm 2.95$ \\
         W$_\text{20}/2$ & $1.11 \pm 0.05$ & V$_\text{max,mid}$ & $-13.48 \pm 3.25$ \\
         W$_\text{50}/2$ & $0.86 \pm 0.07$ & V$_\text{max,mid}$ & $-10.61 \pm 4.47$ \\
         \hline
         \hline
          V$_\mathrm{max,mid}$ & $1.05 \pm0.01$ & V$_\mathrm{max,circ}$ & $-1.61 \pm 0.40$ \\
         \hline
    \end{tabular}
    \caption{Best-fit linear equations $(y = mx+b)$, relating each \hi velocity method ($y$) to the halo speeds ($x$): V$_\text{max,circ}$ or V$_\text{max,mid}$. In the bottom row, we show the relation between V$_\text{max,mid}$ and V$_\text{max,circ}$. }
    \label{VmaxcircEQ}
\end{table}

\bsp	
\label{lastpage}
\end{document}